\documentclass[prb,twocolumn,prb,showpacs,aps]{revtex4}
\usepackage{amsmath,amsfonts,bm,graphicx,dsfont}
\usepackage{epsfig}
\usepackage{amssymb}

\newcommand{\e}{\text{e}}
\newcommand{\be}{\begin{equation}}
\newcommand{\ee}{\end{equation}}
\renewcommand{\i}{\text{i}}
\renewcommand{\d}{\text{d}}

\usepackage{color}
\newcommand{\hlA}[1]{{\color{black}{#1}}}   

\begin{document}

\title{Theory of STM junctions for $\pi$-conjugated molecules on thin insulating films}
\author{Sandra Sobczyk}
\author{Andrea Donarini}
\email{andrea.donarini@physik.uni-r.de}
\author{Milena Grifoni}

\affiliation{Institut f\"ur Theoretische Physik, Universit\"at Regensburg, 93040 Regensburg, Germany}
\date{\today}

\begin{abstract}
A microscopic theory of the transport in a scanning tunnelling microscope (STM) set-up is introduced for $\pi$-conjugated molecules on insulating films, based on the density matrix formalism. A key role is played in the theory by the energy dependent tunnelling rates which account for the coupling of the molecule to the tip and to the substrate. In particular, we analyze how the geometrical differences between the localized tip and extended substrate are encoded in the tunnelling rate and influence the transport characteristics. Finally, using benzene as an example of a planar, rotationally symmetric molecule, we calculate the STM current voltage characteristics and current maps and analyze them in terms of few relevant angular momentum channels.
\end{abstract}


\maketitle
\section{Introduction}
Scanning tunnelling microscopy (STM) is an important tool for imaging surface structures and for studying the electronic properties of individual molecules since its introduction by Binnig and Rohrer \cite{Binnig, Binnig2}.
Various authors have developed theories of STM \cite{Tersoff, Tersoff2, Chen, Garcia, Garcia2, Doyen, Stoll, Buker, Baratoff, Noguera, Calleja_04, Toher}, among  those the famous ones published by Tersoff and Hamann \cite{Tersoff, Tersoff2, Tersoff_rev} in the 1980s.
Their work is the basic theory used to explain STM images without atomic resolution \cite{Chen_book}, i.e. STM images with characteristic feature sizes of $\geq 1$nm, for example the scattered waves of surface states, as well as adsorbates, defects and substitution atoms on the surface \cite{Nanayakkara}. Tersoff and Hamann showed that those experiments, as those on reconstructed Au surfaces, may have a simple explanation. In their articles the tip was modeled as a spherical potential well of radius $R=9 \mathring{A}$, taking the $s$-wave solution of the macroscopic Schr\"odinger equation to describe the electronic tip-state. With Bardeen's perturbation theory of tunnelling \cite{Bardeen}, they showed that the STM image is approximately the Fermi-level local density of states (LDOS) contour of the sample at the center of the sphere. Though the Tersoff-Hamann approach cannot be used to
explain famous STM experiments that show atomic resolution, because it ignores the detailed structure of the tip wave functions. For true atomic resolution, for which the length scale is much smaller than one nanometer, the convolution of tip states and sample states must be taken into account \cite{Hofer}. Chen presented an extension of the Tersoff-Hamann theory that implies more detailed tip-models and allows to interpret higher resolution STM images
\cite{Chen, Chen2, Chen3}. Several other authors suggested that atomic resolution demands small tip-sample distances \cite{Herz, Baratoff, Noguera}, which are not fully described within the Bardeen tunnelling theory \cite{Bardeen}.

In fact the majority of the STM studies of single molecules, in experiment and in theory, has so far been limited to molecules on metals or semiconductors. In these cases the electronic properties of an individual molecule are strongly perturbed by the presence of the substrate electrons. In order to understand the electronic properties of an individual molecule, an electronic decoupling from the supporting substrate is desirable. Hence, in the seminal experiments  [\onlinecite{Repp, Repp2}], STM measurements have been performed on molecules on insulating films having a thickness of only few atomic layers. The layer is in turn grown on top of a metallic substrate. This set-up  allows  to electronically decouple the molecule from the metallic surface, so that electronic properties of individual molecules can be studied. At the same time the electrons can still tunnel through the  insulating
films, facilitating imaging with the low-temperature STM at a low tunnelling current.

In this work we present an STM theory that enables to study the transport properties of individual $\pi$-conjugated molecules in the latter STM configuration. We model the device with a double-barrier tunnelling set-up, and treat its dynamics in the sequential tunnelling limit via a density matrix approach. We show that the geometrical aspects in the coupling to the substrate and the tip, results into significantly different, energy dependent tunnelling rates. Using benzene as an example, we calculate current voltage characteristics and constant height current maps for different biases and substrate work functions, thus simulating STM images with atomic resolution. Due to the rotational symmetry of the benzene molecule we express the theory in the angular momentum basis, and we prove that the tunnelling dynamics from/to the extended substrate  is described by angular momentum channels. Vice versa, the localized tip mixes, in the tunnelling events, the angular momentum states of the molecule. This mixing produces, for specific substrate work functions, negative differential conductance and current blocking also detectable in the topography of the STM surface plots.

\hlA{Both the Pauli and the generalized master equation have been repeatedly used in the modelling of STM junctions \cite{GaoPL_92,StokbroTSQYPG_98,MaG_00,WongG_02,RyndykACR_08,SantandreaGSJ_11}.
Nevertheless, to our knowledge, STM junctions with a thin insulating layer have not been systematically studied within the framework of the generalized master equation.}

This paper is outlined as follows: in section II we present a general transport theory for $\pi$-conjugated molecules in the STM set-up. We introduce the model Hamiltonian of the system and provide a detailed analysis of the tunnelling dynamics in terms of energy dependent tunnelling rates. In section III we apply the theory to a benzene molecule.  The corresponding current voltage characteristics and current maps are discussed in section IV. Finally, conclusions and remarks are presented in section V.

\section{Low energy theory of STM on insulating layers}

\subsection{Hamiltonian and tunnelling amplitudes}

\begin{figure}\centering
\includegraphics[width=0.5\textwidth]{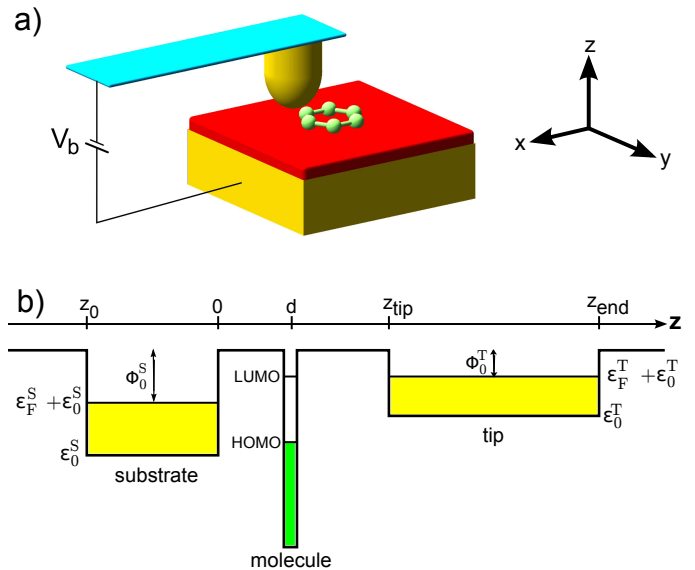}
\caption{(Color online) Panel (a) - Sketch of the investigated STM set-up. A $\pi$-conjugated molecule, here benzene, is separated by a  metal substrate (yellow) only through an ultra-thin insulating film (red). A bias voltage is applied between the substrate and the tip.
Panel (b) - Schematic illustration for the sum of the potentials of the substrate, the molecule and the tip $v = v_{\rm sub} + v_{\rm m} + v_{\rm tip}$, along the $z$ direction. We choose the energy of the vacuum between the molecule and the tip, as well as the energy of the tunnelling barrier between molecule and substrate to be zero. The energies at the bottom of the conduction band of tip and substrate are $\varepsilon_0^{S/T} = -\Phi_0^{S/T}-\varepsilon_F^{S/T}$, where $\varepsilon_F^{S/T}$ are the Fermi energies measured from the band bottom and $\Phi_0^{S/T}$ are the work functions for the tip and the substrate. The work functions are shifted by the applied bias voltage.}
\label{STMandWELL}
\end{figure}

A scanning tunnelling microscopy (STM) set-up with a thin insulating film involves the STM tip, the substrate and the molecule (Fig.~\ref{STMandWELL}a), weakly coupled to each other.
Therefore we can describe the whole system by the total Hamiltonian
\begin{equation}\label{Ham}
 H=H_{\rm m}+H_{\rm sub}+H_{\rm tip}+H_{\rm tun}\;.
\end{equation}
The first term gives the Hamiltonian of an arbitrary $\pi$-conjugated molecule. We assume that only the $\pi$-orbitals contribute to transport. Thus, to each atom is assigned only one orbital (the $2p_z$ orbital orthogonal to the plane of the molecule), while the entire $\sigma$ backbone is included only via the parametrization of the Hamiltonian for the $\pi$-conjugated electrons. The latter, written in the atomic basis, is a simplified version of the Pariser-Parr-Pople (PPP) Hamiltonian \cite{Pariser,Pople}, expressed in terms of the non-interacting H\"uckel-Hamiltonian \cite{Huckel} and a constant interaction term:
\begin{equation}\label{Hmolfull}
\begin{aligned}
 H_{\rm m}= & \sum_{\alpha\sigma}a_{\alpha} d^\dagger_{\alpha\sigma}d_{\alpha\sigma} + \sum_{\alpha \neq \beta\sigma} b_{\alpha \beta} d_{\alpha\sigma}^\dagger
 d_{\beta\sigma}
+\\
&+\frac{1}{2}U\left(N-N_0\right)^2,
\end{aligned}
\end{equation}
where $d_{\alpha\sigma}^\dagger$ creates an electron of spin $\sigma$ in the $p_z$-orbital of the atom $\alpha$, and $\alpha=1,...,M$ runs over the $M$ atoms of the molecule. The hopping energies $b_{\alpha \beta}$ are assigned using the Slater-Koster method \cite{SlaterK}  with atomic parameters and geometrical configurations obtained from the literature. The on-site energy for the atom $\alpha$ is denoted by $a_\alpha$ and can also vary from atom to atom. Finally, the constant interaction model \cite{Kouwenhoven} assumes that the Coulomb interaction between the electrons is parameterized by a constant capacitance $C$, what is finally defining the Coulomb interaction $U=\frac{e^2}{2C}$, where $e$ is the charge quantum. This model also assumes that the discrete single-particle energy spectrum is unaffected by the interactions. Finally, $N = \sum_{\alpha \sigma} d^\dagger_{\alpha\sigma}d_{\alpha\sigma}$ counts the number of $\pi$-electrons in the molecule which is $N_0$ for the neutral case.

The simplicity of the Hamiltonian for the molecule presented here  allows to carry out most of the calculations (specifically the ones relative to benzene presented in sections III and IV) at an analytic level since the many-body eigenstates of the interacting Hamiltonian coincide, in this case, with the ones of the non interacting one. Nevertheless, the transport theory is not affected by the particular choice of the Hamiltonian for the molecule and the transport characteristics remain qualitatively the same for the different models, as far as the symmetry of the states is preserved.

We consider the tip and and the substrate as reservoirs of non interacting electrons. In particular, we describe the metallic substrate as a potential well (see Fig.~\ref{STMandWELL}b) with no confinement in the $x$ and $y$ direction. The associated Hamiltonian $H_{\rm sub}$ reads
\be
 H_{\rm sub}=\sum_{\vec{k}\sigma}\varepsilon^{S}_{\vec{k}}
 c_{S\vec{k}\sigma}^\dagger c_{S\vec{k}\sigma},
\ee
where $\varepsilon^{S}_{\vec{k}} = \varepsilon_0^S + \frac{\hbar^2|\vec{k}|^2}{2m}$ with $\vec{k} = (k_x,k_y,k_z)$ and $c_{S\vec{k}\sigma}^\dagger$ creates an electron of momentum $\vec{k}$ and spin $\sigma$ in the substrate and $|z_0|$ is the $z$ extension of the substrate (see Fig.~\ref{STMandWELL}). The continuous choice also for the $z$ component of the momentum is justified in the limit $|z_0| \gg \lambda_F$ where $\lambda_F$ is the Fermi wave length of the substrate. Only bound states ($\varepsilon^{S}_{\vec{k}}<0$) are considered in the calculation and their explicit wave function is given in the Appendix \ref{APPoverlap}.

An analogous shallow square potential for the $z$ direction  describes the metallic tip. A parabolic confinement in the $x$ and $y$ direction is though added to the model to simulate the spatial localization of the tip states. The tip Hamiltonian reads:
\be
\label{eq:H_tip}
 H_{\rm tip}=\sum_{k_z\sigma}\varepsilon^T_{k_z}
 c_{Tk_z\sigma}^\dagger c_{Tk_z\sigma}\;,
\ee
where $\varepsilon^T_{k_z} = \varepsilon_0^T + \hbar\omega +   \frac{\hbar^2k_z^2}{2m}$ and $c_{Tk_z\sigma}^\dagger$ creates an electron with momentum $k_z$, spin $\sigma$, and in the ground state with respect to the lateral confinement.

We are confident that the particular choice of the confinement for the tip Hamiltonian is not crucial for the results. Nevertheless, as it has already been theoretically predicted \cite{Chen} and experimentally observed \cite{Meyer_prl_2011}, the symmetry of the tip is very important. We will restrict in this work to tip wave functions which are rotationally invariant with respect to an axis perpendicular to the surface of the substrate.

The last term of Eq.~(\ref{Ham}) is the tunnelling Hamiltonian. It contains two parts: one for the substrate-molecule tunnelling, the other for the tip-molecule tunnelling:
\be\label{ht2}
H_{\rm tun}=\sum_{\chi k i\sigma}t_{k i}^\chi c_{\chi k \sigma}^\dagger d_{i\sigma}+h.c.\;.
\ee
The index $i$ denotes the molecular orbital, i.e. the linear combination of the atomic $p_z$ orbitals introduced in Eq.~\eqref{Hmolfull}, $\chi = S,T$ indicates the substrate or the tip and we have introduced the general label $k$ indicating the orbital quantum numbers of both the leads with the identification $k = \vec{k}$ for the substrate and $k = k_z$ for the tip. The coefficient $t_{k i}^{\chi}$ is the tunnelling amplitude that contains all the geometrical information about the tunnelling processes. Denoting by $h=\frac{p^2}{2m}+v_{\rm m}+v_{\rm sub} + v_{\rm tip}$ the single particle Hamiltonian for an electron in the STM set-up, we define this amplitude by
\be
t_{ki}^\chi := \langle \chi k\sigma| h |i \sigma \rangle \;,
\ee
where $|\chi k\sigma\rangle$ and $|i \sigma \rangle$ are eigenstates of the reservoir $\chi$ and of the molecule, respectively. The kinetic energy of the electron is given by $\frac{p^2}{2m}$. The molecule, tip and substrate potentials are denoted by $v_{\rm m}$, $v_{\rm tip}$ and $v_{\rm sub}$, respectively. The $z$-dependence of the total potential $v =v_{\rm m}+ v_{\rm sub} + v_{\rm tip}  $ is schematically shown in Fig.~\ref{STMandWELL}b. It is the sum of three potential wells, for the substrate, molecule and tip where $\varepsilon_0^{\chi}<0$ defines the bottom of the conduction band and $\varepsilon_0^{\chi}+\varepsilon_F^{\chi}<0$ are the Fermi energies. For the tunnelling amplitudes, it follows:
\be\label{amplitude}
\begin{aligned}
t_{k i}^\chi &= \langle \chi k\sigma|\underbrace{ \frac{p^2}{2m}+v_{\rm m}}_{=h_{\rm mol}} |i \sigma \rangle +
 \underbrace{\langle \chi k \sigma|v_{\rm sub}+v_{\rm tip} |i \sigma \rangle}_{\sim0}\\
&=\varepsilon_i\langle \chi k \sigma |i \sigma \rangle=\varepsilon_i\sum_\alpha\langle \chi k \sigma |\alpha \sigma\rangle\langle \alpha\sigma|i \sigma \rangle\;,
\end{aligned}
\ee
where $h_{\rm mol}$ is the non-interacting single-particle H\"uckel-Hamiltonian that satisfies the eigenvalue equation
$h_{\rm mol} |i\sigma\rangle = \varepsilon_i|i\sigma\rangle$.

\hlA{The key observation to understand why the matrix element $\langle \chi k\sigma|v_{\rm sub}+v_{\rm tip} |i \sigma \rangle$ can be neglected while the contribution $\langle\chi k\sigma|v_{\rm m}|i \sigma \rangle$ containing the molecular potential should be retained is the larger penetration length of the lead wave function, with respect to that of the molecular orbital, into the barrier region separating the lead and the molecule.} This difference implies in fact that the relevant integration region for the matrix element $\langle \chi k\sigma|v_{\rm sub} + v_{\rm tip} + v_{\rm m}|i \sigma \rangle$ is shifted towards the molecule. Consequently the kinetic energy contribution should be complemented by the one of the molecular potential. For systems characterized by  states with comparable penetration lengths instead, the relevant integration region is in the tunnelling barrier and the kinetic energy yields the dominant contribution.

The different penetration lengths for the lead and molecule wave functions is justified as follows. First, the spatial extension of the valence orbitals is larger for the metallic atoms of the lead  than for the ones in the conjugated molecule. Moreover, the states in the lead which dominate the tunnelling have no nodal planes perpendicular to the molecular surface (low $k_{\parallel}$) while the HOMO and LUMO states of a conjugated molecule have usually several nodal planes perpendicular to the plane of the molecule. These perpendicular nodal planes are associated to a destructive interference between the atomic wave functions which implies that  the higher the number of nodal planes, the shorter is the extension of the molecular orbital in the direction perpendicular to the molecular plane.

Notice that the energy of the vacuum between the molecule and the tip has been set to zero. Likewise we also set to zero the top of the tunnelling barrier between the molecule and substrate, corresponding to the thin insulating layer. The theory is not affected though by a different value of the potential in the barrier regions as far as the latter is spatially uniform if compared with the product of the lead and molecule wave functions in the same region. A more precise description of the lead potential would in first approximation just lead to a renormalization in Eq.~\eqref{amplitude} of the orbital energy $\varepsilon_i$.

In the last step of Eq.~(\ref{amplitude}) we added the completeness $1=\sum_\alpha|\alpha \sigma\rangle\langle \alpha\sigma|$, where $|\alpha \sigma\rangle$ is the $p_z$-state of the atom $\alpha$,  thus showing that the wanted matrix element can be expressed in terms of the overlap $\langle \chi k \sigma |\alpha \sigma\rangle$ of the lead and the $p_z$-orbital and the basis transformation $\langle \alpha\sigma|i \sigma \rangle$ from the molecular to the atomic orbital. Finally, we obtain for the tunnelling amplitudes:
\be\label{eq:ampliTIP}
t^T_{ki} = \varepsilon_i\sum_\alpha O_T(k_z,\vec{R}_{\rm tip}-\vec{R}_{\alpha}) \langle\alpha\sigma|i\sigma\rangle\;,
\ee
\be\label{ampliSUB}
t^S_{ki} = \varepsilon_{i}\sum_\alpha\e^{-\i\vec{k}_{||}\cdot\vec{R}_\alpha}O_S(\vec{k})\langle\alpha\sigma|i\sigma\rangle\;,
\ee
where $\vec{R}_{\alpha}$ and $\vec{R}_{\rm  tip}$ are the position of the atom $\alpha$ and of the tip, respectively. The overlaps $O_{\chi}$ are given explicitly in the Appendix \ref{APPoverlap} and are calculated using the $p_z$-orbital \cite{Hehre}:
\be
p_z(\vec{r}-\vec{R}_\alpha)= \langle \vec{r}|\alpha_G \rangle = n_G\sum_i \beta_i\, (\vec{r}-\vec{R}_\alpha)\cdot \hat{e}_z \,\e^{-\alpha_i |\vec{r}-\vec{R}_{\alpha}|^2}\;,
\ee
where $\hat{e}_z$ is the versor in the direction perpendicular to the molecular plane, the coefficient $n_G$ assures  normalization and the parameters $\alpha_i$ and $\beta_i$, that we show in table \ref{tab_alpha_beta} for the specific case of a carbon atom, define the gaussian representation for a Slater type orbital commonly used in DFT calculations \cite{Jensen, NWchem}. Analogous parametrizations are available also for other atoms and allow a straightforward application of the model to generic planar $\pi$-conjugated molecules.
\begin{table}
 \caption{Parameters $\alpha_i$, $\beta_i$ used for the Gaussian $p_z$-orbitals}
 \begin{tabular}{l||c|c|c}
 $i$& $1$& $2$& $3$\\
      \hline\hline
 $\alpha_i \bigl[\frac{1}{\mathring{A}^2}\bigr]$  & $0.368$ & $ 1.113$ & $4.997$ \\ \hline
 $\beta_i \bigl[\frac{1}{\mathring{A}^{5/2}}\bigr] $ & $0.502$ & $1.438$ & $2.620$
 \end{tabular}
 \label{tab_alpha_beta}
 \end{table}
The overlap functions of the substrate and the tip are qualitatively different since they reflect the different geometries of the corresponding contacts. The plane wave description of the electrons in the substrate implies that in Eq.~(\ref{ampliSUB}) the position of the atom $\vec{R}_\alpha$ only appears in the phase factor as a scalar product with the component of the momentum parallel to the substrate, $\vec{k}_{||}$.
Additionally we obtain a function that only depends on the electron's momentum $\vec{k}$ in the substrate and on the thickness of the insulating barrier. This particular form already suggests that the tunnelling between the substrate and the molecule is not an incoherent collection of tunnelling events happening in correspondence to the different atoms since their position is recorded in the phase of the tunnelling amplitude. Some of the consequences of this spatial coherence will appear more clearly in section \ref{sec:Benzene} where we analyze the special case of a benzene STM junction. The overlap function for the tip is more complex. Due to the cylindrical symmetry of the tip and atomic orbital with respect of their rotational axes, we can only further conclude that only the modulus of the component of $\vec{R}_{\rm tip}-\vec{R}_{\alpha}$ parallel to the molecular plane influences the tunnelling (see Appendix \ref{APPoverlap}).

\subsection{Tunneling dynamics}

Our method of choice to treat the dynamics in the regime of weak coupling between system and leads is the Liouville equation method. A detailed discussion and derivation of the equations of motion for the reduced density operator of the system can be found e.g. in \cite{Blum,Darau}; we will give here only a short overview adapted to the STM set-up.

We start from the Liouville equation for the total density operator $\rho(t)$ of the whole system consisting of the molecule, the tip and the substrate. Using the interaction picture and treating the tunnelling Hamiltonian (\ref{ht2}) as a perturbation we get:
\be\label{rhodt}
\i \hbar \frac{\d \rho^I(t)}{\d t}=[H_{\rm tun}^I(t),\rho^I(t)]\;,
\ee
where the subscript $I$  indicates the use of the interaction picture. Since we are not interested in the microscopic state of the leads, we focus on the time evolution of the reduced density matrix (RDM) $\sigma={\rm Tr}_{S+T}\{\rho(t)\}$, which is formally obtained by taking the trace over the unobserved degrees of freedom of the  tip and the substrate. The equation of motion for the RDM reads to lowest non-vanishing order in the coupling to the substrate and the tip \cite{Donarini}
\be\label{GME}
\dot{\sigma} = -\frac{\i}{\hbar}[H_{\rm m},\sigma]-\frac{\i}{\hbar}[H_{\rm eff},\sigma]+\mathcal{L}_{\rm tun}\sigma\ := \mathcal{L}\sigma.
\ee
The first term of this so called generalized master equation (GME) gives the coherent evolution of the system in absence of the substrate and the tip. In the secular approximation we only keep coherences between degenerate states and thus this term vanishes \cite{Blum}. The commutator with $H_{\rm eff}$ includes the normalization of the coherent dynamics introduced by the couplings to the leads. Finally, the operator $\mathcal{L}_{\rm tun}$ describes the sequential tunnelling processes. The sum of these three contributions defines the Liouville operator $\mathcal{L}$.

Let us concentrate first on the tunnelling processes occurring in the system. The corresponding contribution to the master equation, projected into the subspace of $N$-particles and energy $E$ reads:
\begin{widetext}
\be\label{GMEshort}
\begin{split}
\mathcal{L}_{\rm tun}\sigma^{NE} = & -\frac{1}{2}
\sum_{\chi \tau} \sum_{ ij} \left\{
\mathcal{P}_{NE}
\left[
d^\dagger_{i\tau}
\Gamma_{ij}^{\chi}(E-H_{\rm m})f^-_{\chi}(E-H_{\rm m})
d_{j\tau} +
d_{j\tau}
\Gamma^{\chi}_{ij}(H_{\rm m}-E)f^+_{\chi}(H_{\rm m}-E)
d^\dagger_{i\tau}
\right]
\sigma^{NE} + h.c.
\right\}\\
& + \sum_{\chi \tau} \sum_{ij E'}
\mathcal{P}_{NE}
\left[
d^\dagger_{i\tau}
\Gamma_{ij}^\chi(E-E')\sigma^{N-1E'}f^+_{\chi}(E-E')
d_{j\tau} +
d_{j\tau}
\Gamma_{ij}^\chi(E'-E)\sigma^{N+1E'}f^-_{\chi}(E'-E)
d^\dagger_{i\tau}
\right]\mathcal{P}_{NE}
\end{split}
\ee
\end{widetext}
where $\sigma^{NE}:= \mathcal{P}_{NE}\sigma \mathcal{P}_{NE}$ being $\mathcal{P}_{N\!E} :=\sum_{l}|NEl\rangle\langle NEl|$ the projection operator on the subspace of $N$ particles and energy $E$, and $l$ the additional quantum number that distinguishes between degenerate states. Moreover, $f_\chi^+(x)$ is the Fermi function for the lead $\chi$, $f_\chi^+(x):=f(x-\mu_\chi)$, and $f_\chi^-(x):=1-f_\chi^+(x)$. The terms proportional to $f_\chi^+(x)$ describe in \eqref{GMEshort} tunnelling events \emph{to} the molecule, while the tunnelling \emph{out} of the molecule is associated to $f_\chi^-(x)$. Finally $\mu_{\chi}$ stands for the electro-chemical potentials of the substrate or the tip. They are defined via the applied bias voltage as $\mu_S =\mu_0+(1-c)eV_b$,  $\mu_T =\mu_0 -c\,eV_b$ and consequently $ eV_b = \mu_S-\mu_T\;,$ with the electron charge $e$, the equilibrium potential $\mu_0$ and the coefficient $c$ governing the relative bias drop at the tip and the substrate. A symmetrical potential drop is obtained for $c = 1/2$, while for $c = 1$ the bias drops completely at the tip-molecule interface. Finally $\mu_0 = -\Phi_0$ relates the equilibrium chemical potential to the work function and, in equilibrium, the work functions of the two leads are assumed equal. Beside the Fermi function, the tunnelling rates are characterized by the geometrical component:
\be\label{rateSubTip}
\Gamma_{ij}^\chi(\Delta E) = \frac{2\pi}{\hbar}
\sum_{k}
\left(t^\chi_{ki}\right)^*
t^\chi_{kj}\,
\delta (\varepsilon^{\chi}_k-\Delta E)\;.
\ee
The argument $\Delta E$ of the rate $\Gamma_{ij}^\chi$ is the energy difference $E_{N+1} - E_{N}$ of the many body states involved in the tunnelling process, sometimes written in Eq.~\eqref{GMEshort} in terms of the operator $H_{\rm m}$. \hlA{Notice that the rate $\Gamma_{ij}^\chi$ vanishes if $\Delta E > 0$ since we restrict the Hilbert space of the leads to the bound states i.e. $\varepsilon_k < 0$.} The quantity $\Gamma_{ij}^\chi$ plays a central role in the theory and in the following section we will discuss its  calculation in detail for the tip and the substrate case using the example of a benzene molecule.

A natural expression for the current operators is obtained in terms of the time derivative of the reduced density matrix:
\be\label{current1}
\langle I_{\rm sub} + I_{\rm tip}\rangle= \sum_{NE}{\rm Tr}\left\{ N\dot{\sigma}^{NE}\right\}\;,
\ee
where $I_{\rm sub/tip}$ are the current operators calculated for the substrate and the tip interfaces. Conventionally we assume the current to be positive when it increases the charge on the molecule. Thus, in the stationary limit, $\langle I_{\rm sub} + I_{\rm tip}\rangle$ is zero. The stationary current is obtained as the average:
\begin{equation}\label{current2}
 \langle I_{\rm sub} \rangle= {\rm Tr}\left\{ \sigma_{\rm stat}I_{\rm sub}\right\}
 =-\langle I_{\rm tip}\rangle\;,
\end{equation}
 where $\sigma_{\rm stat}=\lim_{t\to\infty}\sigma(t)$ is the stationary density operator that can be found from
\begin{equation}\label{null}
 \dot{\sigma}_{\rm stat}=\mathcal{L}\sigma_{\rm stat}=0\;,
\end{equation}
where $\mathcal{L}$ is the Liouville operator. Finally, by following exactly the procedure given in [\onlinecite{Darau}], we find the explicit expressions for the current operators:
\begin{equation}
\label{subcurr}
\begin{split}
 I_\chi=\sum_{NE\sigma ij}
 \mathcal{P}_{NE}
 \biggr[
 d_{j\sigma}\Gamma_{ij}^\chi(H_{\rm m}-E) f_\chi^+(H_{\rm m}-E)d_{i\sigma}^\dagger\\
 -d_{i\sigma}^\dagger \Gamma_{ij}^\chi(E-H_{\rm m}) f_\chi^-(E-H_{\rm m})d_{j\sigma}
 \biggl]
 \mathcal{P}_{NE}\;,
\end{split}
\end{equation}
 where the energy renormalization terms, present in the GME, do not appear.

Since the tunnelling changes the number of electrons on the molecule, the latter behaves as an open system and it is useful to introduce the operator $H'_m = H_{\rm m} - \mu_0N$ where $N$ counts the number of electrons on the molecule. For example, at zero temperature and zero bias the equilibrium is reached when the molecule is in the ground state of $H'_m$ and not of $H_{\rm m}$. As we have already shown elsewhere \cite{Donarini2}, also the non-equilibrium conditions for transport can be better understood in terms of the spectrum of $H'_m$. For this reason in Figs. \ref{Rate} and \ref{TipTRate} the geometrical part of the rates is plotted as a function of $\Delta E':= \Delta E - \mu_0$.

\section{Theory applied to benzene}
\label{sec:Benzene}

\begin{figure}\centering
\includegraphics[width=0.3\textwidth]{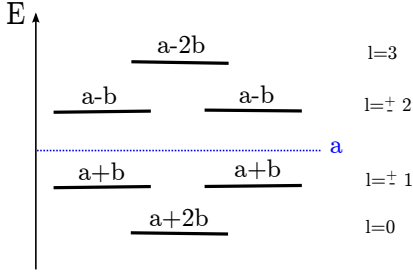}
\caption{Energy levels of the H\"uckel Hamiltonian and the corresponding values of the angular momentum $l$.}
\label{huckelenergy}
\end{figure}

The molecular orbitals of benzene are also eigenfunctions of the projection $l$ of the angular momentum along the main rotational axis, which we assume to be the $z$-axis. Therefore, the basis transformation that occurs in Eq.~(\ref{amplitude}) reads for a benzene molecule
\begin{equation}
\label{eq:basissub}
\langle\alpha\sigma|l\sigma\rangle =\frac{1}{\sqrt{6}}\e^{\i\frac{2\pi}{6}\alpha l}\;
\end{equation}
and the corresponding single particle eigenenergies $\varepsilon_l$, occurring in the Eqs.~(\ref{eq:ampliTIP}) and (\ref{ampliSUB}) for
the tunnelling amplitudes, read:
\be
\varepsilon_l=a+2b\cos\left(\frac{2\pi}{6}l\right)\;.
\ee
For a benzene molecule the possible values of the angular momentum quantum number $l$ are $0\, ,\pm 1\, ,\pm 2\, ,3$ corresponding to the energy level scheme of the H\"uckel Hamiltonian shown in Fig.~\ref{huckelenergy}. Since the Hamiltonian is invariant under the discrete rotations of angles $n\pi/3$ with $n \in \mathbb{Z}$, the same quantum numbers also label the many-body eigenstates of the benzene molecule, irrespective of the complexity of the description of the Coulomb interaction \cite{Darau}. All the single particle states show a twofold spin degeneracy but only few states possess an additional twofold orbital degeneracy. The latter is essential for the explanation of the transport features of benzene within an STM experiment.

\subsection{The substrate-molecule tunnelling rates}

\begin{figure}\centering
\includegraphics[width=0.4\textwidth]{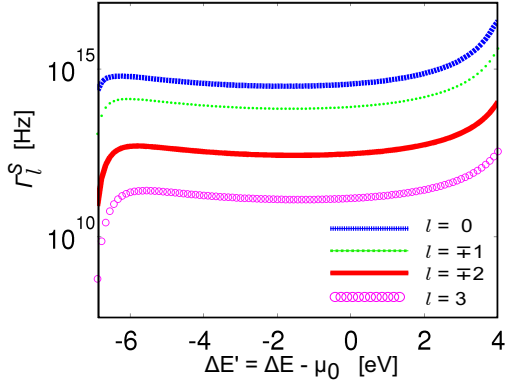}
\caption{(Color online) Tunneling rate $\Gamma_l^S$  describing substrate-molecule tunnelling processes for different angular momentum quantum numbers $l$. \hlA{The thickness of the substrate barrier is $d = 3$\AA, while work function and Fermi energy are respectively $\phi_0^S = 4 eV$ and $\varepsilon_F^S = 7 eV$}}
\label{Rate}
\end{figure}

Let us start with a detailed discussion of the substrate-molecule tunnelling rate. To perform  the sum over the momenta $\vec{k}$ in Eq.~(\ref{rateSubTip}) we transform it into energy integrals, using the definitions $\varepsilon_{||}:= \tfrac{\hbar^2|\vec{k}_{||}|^2}{2m}$ and $\varepsilon_z:= \tfrac{\hbar^2 k_z^2}{2m}$:
\begin{gather}
\sum_{\vec{k}}=\sum_{\vec{k}_{||}}\sum_{k_z}\;,\\
\label{integralkp}
\sum_{\vec{k}_{||}}\rightarrow S\frac{m}{\hbar^2}\int_0^{2\pi}\d\vartheta \int_0^{\varepsilon_F^S+\Phi_0^S}\d \varepsilon_{||}\;,\\
\sum_{k_{z}}\rightarrow|z_0|\frac{1}{\hbar}\sqrt{\frac{m}{2}} \int_0^{\varepsilon_F^S+\Phi_0^S}\d \varepsilon_{z}\frac{1}{\sqrt{\varepsilon_z}}\;,
\end{gather}
where the volume $V = |z_0|S$ is canceled out in the thermodynamic limit by the normalization of the orbitals which define the overlap function. Moreover we observe that Eq.~(\ref{rateSubTip}) requires the calculation of the product
\begin{equation}\label{ttS1}
\begin{aligned}
&\left(t_{\vec{k}l}^S\right)^*t_{\vec{k}l'}^S=\\
&\varepsilon_l\varepsilon_{l'}
\sum_{\alpha\alpha'}
|O_S(\vec{k})|^2
\langle\alpha\sigma|l\sigma\rangle
\langle\alpha'\sigma|l'\sigma\rangle
\e^{+\i\vec{k}_{||}\cdot(\vec{R}_\alpha-\vec{R}_{\alpha'})}\;.
\end{aligned}
\end{equation}
We write the exponential function in Eq.~(\ref{ttS1}) as
 $\e^{+\i\vec{k}_{||}\cdot(\vec{R}_\alpha-\vec{R}_{\alpha'})} =  \e^{+\i|\vec{k}_{||}||\vec{R}_\alpha-\vec{R}_{\alpha'}|\cos\vartheta}$ and the equation finally becomes

\be\label{ttS}
\begin{aligned}
\left(t_{\vec{k}l}^S\right)^*t_{\vec{k}l'}^S&=
\frac{1}{6}\varepsilon_l\varepsilon_{l'}\sum_\gamma\e^{-\i\frac{2\pi}{6}l'\gamma}\sum_{\alpha}
\e^{-\i\frac{2\pi}{6}\alpha(l-l')}\\
&\times\e^{+\i\sqrt{\frac{2m}{\hbar^2}\varepsilon_{||}}|\Delta\vec{R}_\gamma|\cos\vartheta}|O_S(\varepsilon_{||},\varepsilon_z)|^2\;,
\end{aligned}
\ee
where we introduced $\alpha-\alpha':=\gamma\;,\; |\vec{R}_\alpha-\vec{R}_{\alpha'}|=|\Delta \vec{R}_\gamma|$.
We insert Eq.~(\ref{ttS}) in the substrate case of  Eq.~(\ref{rateSubTip}) and, after solving the integral over $\d \vartheta$, we find:
\be\label{GMEshortNEW2}
\begin {aligned}
&\Gamma_{ll'}^S(\Delta E)=
\frac{\pi^2}{6\hbar^4}m^\frac{3}{2}\sqrt{2}\varepsilon_l\varepsilon_{l'}\sum_{\alpha}
\e^{+\i\frac{2\pi}{6}\alpha(l-l')}\\
&\times\!\int_0^{\varepsilon_F+\Phi_0}\!\d\varepsilon_{||}\,\int_0^{\varepsilon_F+\Phi_0}\!\d\varepsilon_z\,\frac{V}{\sqrt{\varepsilon_z}}
\sum_\gamma J_0\left(\sqrt{\frac{2m}{\hbar^2}\varepsilon_{||}|}\Delta\vec{R}_\gamma|\right)\\
&\times|O_S(\varepsilon_{||},\varepsilon_z)|^2
 \e^{+\i\frac{2\pi}{6}l'\gamma}\delta(\varepsilon_k-\Delta E)\;,
\end{aligned}
\ee
with $J_0(x)$ the zero-order Bessel function. Finally, using the relation
\be\label{deltall}
\sum_\alpha\e^{\pm\i\frac{2\pi}{6}\alpha(l-l')}=6\,\delta_{ll'}\;,
\ee
and the fact that $\sum_\gamma\e^{\i\frac{2\pi}{6}l\gamma}=\sum_\gamma\e^{-\i\frac{2\pi}{6}l\gamma}$  the integral over $\varepsilon_{||}$ yields
\be
\begin{aligned}\label{Trate}
&\Gamma_{ll'}^S(\Delta E) =  \delta_{ll'}
\frac{\pi^2}{\hbar^4}m^\frac{3}{2}\sqrt{2}
\,\varepsilon_l^2
\int_0^{\varepsilon_F^S+\Phi_0^S}\d \varepsilon_z \frac{V}{\sqrt{\varepsilon_z}}\\
&\times
\sum_\gamma
J_0\left(
\sqrt{\frac{2m}{\hbar}\left(\Delta E -\varepsilon_z-\varepsilon_0^S\right)} |\Delta \vec{R}_\gamma|
\right) \e^{-\i\frac{2\pi}{6}l\gamma}\\
&\times |O_S(\Delta E -\varepsilon_z -\varepsilon_0^S,\varepsilon_z)|^2\\
&\times\Theta\left(\Delta E -\varepsilon_z-\varepsilon_0^S\right)
\Theta\left(\varepsilon_z-\Delta E \right)\;.
\end{aligned}
\ee
The integral in Eq.~(\ref{Trate}) has to be solved numerically.
The main result of the latter calculations is
\be
 \Gamma_{ll'}^S(\Delta E)=\delta_ {ll'}\Gamma_{l}^S(\Delta E)\;,
 \label{eq:Sratematrix}
\ee
which ensures that tunnelling processes involving the substrate are happening through \emph{angular momentum channels} because a mixing of angular momenta is not allowed in the substrate. We will see that this only happens for substrate-tunnelling-processes, while there is no conservation rule for angular momenta in the tip-tunnelling case. The function  $\Gamma_l^S(\Delta E)$ is the geometrical rate and we plot it in Fig.~\ref{Rate} for different angular momenta.
The rates decrease of several order of magnitudes by increasing the absolute value of the projection of the angular momentum $l$. This is the direct consequence of the decreasing extension of the molecular orbitals in the direction perpendicular to the molecular plane with increasing the number of vertical nodal planes.

\hlA{The lower limit of the energy axis in Fig.~\ref{Rate} is $-\varepsilon^S_{\rm F}$ while the upper limit is the work function $\phi^S_0$. These limits are set by the substrate model in which only bound states of a single band are taken into account ($\varepsilon^S_0 <\epsilon^S_k < 0$). While approaching the low energy limit $\Delta E = -\varepsilon^S_{\rm F}$ both the density and the penetration length of the states in the substrate which contribute to the rate reduce, hence the turn down. On the other hand, the increasing of the density of states and of the penetration length explains the turn up at the upper energy border ($\Delta E = \phi_0^S$)}.

\subsection{The tip-molecule tunnelling rates}

\begin{figure}\centering
\includegraphics[width=0.4\textwidth]{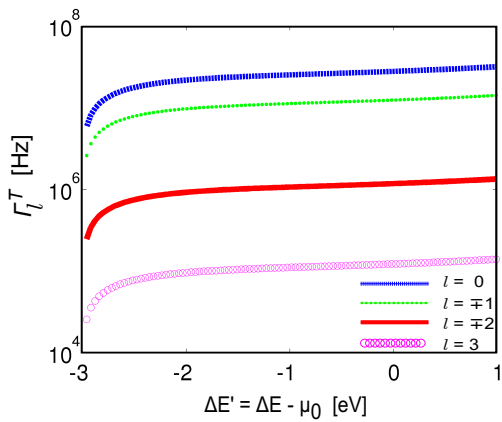}
\caption{(Color online) Diagonal elements $\Gamma_l^T$ of the tip tunnelling rate matrix $\Gamma_{ll'}^T$ for the different angular momentum states. \hlA{The rates are calculated assuming $z_{\rm tip}-d = 3.5$ \AA, $\phi_0^T = 4 eV$, $\varepsilon_F^T = 7 eV$ and $\hbar\omega = 4 eV$. The presence of the harmonic confinement explains also the different energy limits with respect to the ones of Fig.~\ref{Rate}. The lower limit is at $-\varepsilon_F^T + \hbar \omega$ while the upper limit is at $-\varepsilon_F^T + 2\hbar \omega$.}}
\label{TipTRate}
\end{figure}

\begin{figure}\centering
\includegraphics[width=0.4\textwidth]{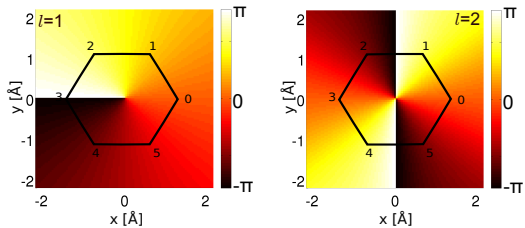}
\caption{(Color online) Phase $\phi_l$ of the tunnelling rate matrix $\Gamma^{T}_{ll'}$, Eq.~\eqref{eq:phasetip}. The phase is almost constant if the tip is moved along the radii outgoing from the center of the molecule. The carbon atoms are labelled by $\alpha = 0,\ldots,5$. }
\label{SymmetryPhase}
\end{figure}

Let us now discuss the tunnelling events happening between the tip and the molecule. To model the tip we consider a harmonic confinement in the $x$ and $y$ directions. By considering the tip to be in the ground state of the $2$-dimensional harmonic oscillator, the longitudinal energy $\varepsilon_{||}$ is fixed to be the
constant $\varepsilon_{||} = \hbar\omega$, cf.~below Eq.~\eqref{eq:H_tip}. The sum in Eq.~(\ref{rateSubTip}) thus transforms into a sum over $k_z$. Because of the relation $k_z = \sqrt{\frac{2m}{\hbar^2}\varepsilon_z}$ we can replace the sum by the integral:
$\sum_{k_{z}}\rightarrow\frac{1}{\hbar}\sqrt{\frac{m}{2}} \int\d \varepsilon_{z}\frac{|z_{\rm end}-z_{\rm tip}|}{\sqrt{\varepsilon_z}}$.
Eq.~(\ref{eq:ampliTIP}) implies
\be
\begin{split}
\left(t^T_{k_z l}\right)^*t^T_{k_z l'} = &
\frac{1}{6}\sum_{\alpha  \alpha'} \varepsilon_l\varepsilon_{l'} \e^{-\i\frac{2\pi}{6}(\alpha l - \alpha'l')}\\
& \times O_T^*(k_z,\vec{R}_{\rm tip}-\vec{R}_\alpha)
O_T(k_z,\vec{R}_{\rm tip}-\vec{R}_{\alpha'})\;,
\end{split}
\ee
that we insert in Eq.~(\ref{rateSubTip}). After solving the energy integral we finally find
\be
\begin{aligned}\label{tunratetip}
 \Gamma&_{ll'}^T(  \Delta E,\vec{R}_{\rm tip}) =
\frac{\pi}{6\hbar^2}\sqrt{\frac{m}{2}}
\sum_{\alpha \alpha'} \varepsilon_l\varepsilon_{l'}\e^{-\i\frac{2\pi}{6}(\alpha l - \alpha'l')}\\
& \times O_T^*(\tilde{k},\vec{R}_{\rm tip}-\vec{R}_\alpha)
O_T(\tilde{k},\vec{R}_{\rm tip}-\vec{R}_{\alpha'})
\frac{|z_{\rm end}-z_{\rm tip}|}
{\sqrt{\Delta E -\varepsilon_0^T - \hbar\omega}}\\
&\times \Theta(\Delta E -\hbar\omega-\varepsilon_0^T)
\Theta(2\hbar\omega-\Delta E +\varepsilon_0^T)\;,
\end{aligned}
\ee
where $\tilde{k} = \sqrt{\frac{2m}{\hbar^2}(\Delta E - \hbar\omega - \varepsilon_0^T)}$. The occurrence of both $l$ and $l'$ in the latter equation, shows that a mixing of angular momenta during the tip-tunnelling process takes place. Upon inspection of Eq.~(\ref{tunratetip}) we find some important relations obeyed by the tunnelling rate, where we use the fact that $l$ and $l'$ always occur in the form $ l' = \pm l$:
\begin{equation}
\label{GammaT_rel}
\begin{split}
 &\Gamma_{ll}^T=\Gamma_{\bar{l}\bar{l}}^T=(\Gamma_{\bar{l}\bar{l}}^T)^*\in \mathbb{R},\\
 &\Gamma_{l\bar{l}}^T=(\Gamma^T_{\bar{l}l})^*,\\
 &|\Gamma_{ll}^T|=|\Gamma_{l\bar{l}}^T|=\Gamma_{ll}^T,
\end{split}
\ee
where we have introduced the notation $\bar{l} \equiv -l$. Thanks to the relations \eqref{GammaT_rel} we can rewrite the tunnelling rate as
\be\label{tiprateshort}
\Gamma_{ll'}^T =
\Gamma_l^T
\e^{-\i \phi_{l}(\vec{R}_{\rm tip})(l-l')/l},
\end{equation}
where $\Gamma_l^T\equiv\Gamma^T_{ll}$, which implies the existence of an angular momentum dependent phase when $l\ne l'$. In Fig.~\ref{TipTRate} we show the diagonal elements of the rate matrix $\Gamma^T_{ll'}$ exemplified for $l = \pm 1$ and $l = \pm 2$. As for the substrate, the channel $l = \pm 1$ leads to a much larger rate than the channel $l=\pm 2$. The phase in the off diagonal elements depends on the tip position $\vec{R}_{\rm tip}$ and it is calculated as

\be\label{eq:phasetip}
\phi_l(\vec{R}_{\rm tip})= \arg(t^T_{k_z l}).
\end{equation}
In Fig.~\ref{SymmetryPhase} we show the values acquired by the phase $\phi_l(\vec{R}_{\rm tip})$ as a function of the tip position. The phase is approximately constant along the radii leaving the center of the molecule. Due to the cylindrical symmetry of the tip wave function a good approximation to the phase $\phi_l(\vec{R}_{\rm tip})$ is given by:

\begin{equation}
\label{eq:Phase_approx}
\phi_l(\vec{R}_{\rm tip}) = l \theta_{\rm tip},
\end{equation}
where $\theta_{\rm tip}$ is the angle describing the projection of the tip position on the molecular plane if the origin is the center of the molecule. By convention we assume $\theta_{\rm tip} = 0$ along the radius that intersects the position of the atom $0$ of the molecule (see Fig.~\ref{SymmetryPhase}). The derivation of this simple expression for $\phi_l$ as well as a discussion on its limits of validity are given in Appendix \ref{app:phase}. Notice that the phase defined in Eq.~\eqref{eq:phasetip} only depends on $\vec{R}_{\rm tip}$ even if $O_\alpha$ contains the bias. Nevertheless, the tunnelling rate Eq.~(\ref{tiprateshort}) depends on the bias via the Fermi energy.

In Fig.~\ref{SymmetryPhase} the position of the $\phi_l = 0$ line is arbitrary and connected to the arbitrary choice of overall phase for the molecular orbital with angular momentum $l$. A different choice of the overall phase would, nevertheless simply appear as a rigid rotation of the plots. Moreover, this arbitrariness has no influence on the current voltage characteristics of the junction.

In the substrate the tunnelling matrix is diagonal and proportional to the identity matrix, independent of the basis representation, see Eq.~\eqref{eq:Sratematrix}. In contrast, according to Eq.~\eqref{tiprateshort}, off-diagonal elements are present in the tip-tunnelling matrix which, in the basis $\{|l\rangle$, $|\bar{l}\rangle \}$, reads
\be\label{tipmatrix1}
{\bf \Gamma}^{T}=\Gamma_{l}^T
\begin{pmatrix}
 1 & \e^{-2 \i \phi_l(\vec{R}_{\rm tip})}\\
 \e^{+ 2\i \phi_l(\vec{R}_{\rm tip})}& 1
\end{pmatrix}.
 \ee

An interesting effect of the localized character of the tunnelling from/to the tip can be better appreciated by switching to the basis which diagonalizes the matrix in Eq.~ (\ref{tipmatrix1}). The substrate rate matrix is still proportional to the identity matrix. For the tip rate matrix we get instead:
\be\label{tipmatrix}
{\bf \Gamma}^T
= \Gamma_{l}^T
\begin{pmatrix}
 2 & 0\\
 0 & 0
\end{pmatrix}\;.
\ee

One diagonal element becomes zero, indicating that there are states which are coupled to the substrate but not to the tip. The decoupled state represents a blocking state, which can be populated by a tunnelling event from (to) the substrate but cannot be depopulated by a tunnelling event to (from) the tip. The presence of blocking states is visible in the current-voltage characteristic, as we will discuss in the next section.

\subsection{Stationary density matrix}

By combining now the expression for the tunnelling rates with the dynamical equation Eq.~\eqref{GMEshort} we can calculate the time evolution of the reduced density matrix associated to $\mathcal{L}_{\rm tun}$ and the corresponding stationary state. The stationary density matrix is block diagonal in particle number, energy and spin. In particular, if we restrict the dynamics to low biases, the only relevant states entering the dynamics are the states $|5_g l \tau \rangle$, $|6_g 0 0 \rangle$, and $|7_g l \tau \rangle$, being the cation, neutral and anion ground states respectively. The neutral ground state is non degenerate while the anion and cation are four times degenerate, due to the combination of the spin and orbital degeneracies. The specific form of the stationary density matrix depends on the bias, the temperature, and the tip position. Nevertheless, due to the form of the tunnelling rate matrices, the two dimensional sub-blocks corresponding to orbitally degenerate states have always the following structure:
\begin{equation}\label{sigmastat}
 \sigma_{\rm stat}^{\bar{N} E_g \tau}=
 \begin{pmatrix}
 A&B\e^{-2\i\phi_l(\vec{R}_{\rm tip})}\\
 B\e^{+2\i\phi_l(\vec{R}_{\rm tip})}& A
 \end{pmatrix}\;,
\end{equation}
where $\bar{N} = 5,7$, the spin $\tau = \uparrow,\downarrow$ and the parameters $A,B$ are functions of the tip position $\vec{R}_{\rm tip}$ and of the bias $V_b$ (see Appendix \ref{AppsigmaStat}). This result is a posteriori not surprising. The comparison of Eq.~\eqref{sigmastat} with Eq.~\eqref{tipmatrix1} reveals in fact that the density matrix and the rate matrices are diagonalized by the same basis transformation (the substrate rate matrix is diagonal in all bases). Thus, the form of $\sigma_{\rm stat}$ could be calculated from the observation that the dynamics of the populations and the coherences is decoupled when expressed in the eigenbasis of the rate matrices. It should be noticed that the diagonalizing basis depends on the phase, see Eq.~\eqref{eq:phasetip}, which in turn depends on the tip position. Thus it is not possible to describe the system using only populations in a unique basis valid for all the positions of the tip.

\subsection{The effective Hamiltonian}
Until now we only concentrated on the sequential tunnelling processes in the system. We still have to discuss the imaginary term in Eq.~\eqref{GME} which contains the effective Hamiltonian $H_{\rm eff}$. The latter is defined as:
\begin{widetext}
\be\label{eq:H_eff}
H_{\rm eff}=
\frac{1}{2\pi} \sum_{NE}\sum_{\chi \sigma} \sum_{ll'}
\mathcal{P}_{NE}\Big[
d_{l\sigma}^\dagger
\Gamma^\chi_{ll'}(E-H_{\rm m})
p_\chi(E-H_{\rm m})
d_{l'\sigma}
+
d_{l'\sigma}
\Gamma^\chi_{ll'}(H_{\rm m}-E)
p_\chi(H_{\rm m}-E)
d_{l\sigma}^\dagger
\Big]\mathcal{P}_{NE},
\ee
\end{widetext}
with the projector $\mathcal{P}_{NE} = \sum_{n} |NE n\rangle \langle NE n|$ and the principal part functions $p_\chi(x)=-{\rm Re}\Psi\left[\frac{1}{2}+\frac{\i}{2\pi k_B T}(x-\mu_\chi)\right]$, with $T$ being the temperature and $\Psi$ the digamma function. Eq.~\eqref{eq:H_eff} shows that the effective Hamiltonian is block diagonal in particle number and energy, exactly as the density matrix in the secular approximation. Consequently, it  only influences the dynamics of the system in presence of degenerate states with corresponding subblocks larger than a mere complex number. For the sake of simplicity we will include in the following calculations only the anion ground states, (i.e. the spin and orbitally degenerate $7$ particle ground states). Analogous arguments holds for all the other degenerate states of the molecule.

If $ \Gamma_{ll'} \propto \delta_{ll'}$  (substrate case, see Eq.~(\ref{eq:Sratematrix})), the effective Hamiltonian $H_{\rm eff}$  in the 7 particle ground state subspace is proportional to the identity matrix, as can be proven from Eq.~\eqref{eq:H_eff} remembering that $H_{\rm m}$ conserves the angular momentum and it is invariant under the symmetry operation that brings $|7_g l \tau \rangle$ into $|7_g \bar{l} \tau \rangle$ and moreover that $\Gamma^S_{ll} = \Gamma^S_{\bar{l}\bar{l}}$. Thus, the substrate contribution to $H_{\rm eff}$ trivially commutes with $\sigma_{\rm stat}$. If the angular momenta $l$ and $l'$ can mix, like in the tip case, $H_{\rm eff}$ acquires off diagonal terms and a more detailed discussion is required. In particular, the form of the off diagonal elements depend on the particular model taken to describe the interaction on the molecule. As shown in the Appendix \ref{APP_omega_L}, within the constant interaction model, the effective Hamiltonian for the tip can be written in the form:
\be
H_{\rm eff}^T = \omega L\;,
\ee
where
\be
\label{omegaHeff}
\begin{aligned}
\omega = \frac{1}{\pi}&
\langle 7_g l\sigma|d_{l\sigma}^\dagger|6_g\,0\,0 \rangle
\langle 6_g\,0\,0|d_{\bar{l}\sigma}|7_g\bar{l}\sigma\rangle\\
&\times \Gamma^T_l(E_{7g}-E_{6g})p_T(E_{7g}-E_{6g})
\\
+ \frac{1}{\pi}&
\langle 7_g l\sigma|d_{\bar{l}\sigma}|8_g\,0\,2\sigma \rangle
\langle 8_g\,0\, 2\sigma|d^{\dagger}_{l\sigma}|7_g\bar{l}\sigma\rangle
\\
& \times\Gamma^T_l(E_{8g}-E_{7g})p_T(E_{8g}-E_{7g})
\end{aligned}
\ee
is the renormalization of the Bohr frequencies for the system and
\be\label{RotOP}
L= \frac{\hbar}{2}
\begin{pmatrix}
 1 & \e^{-2\i\phi_l(\vec{R}_{\rm tip})}\\
\e^{+2\i\phi_l(\vec{R}_{\rm tip})} & 1
\end{pmatrix}\;.
\ee
Hence the effective Hamiltonian $H_{\rm eff}^T$ commutes with the stationary density operator $\sigma_{\rm stat}$ given in Eq.~(\ref{sigmastat}). In conclusion, even if different from zero, the effective Hamiltonian does not contribute to the stationary dynamics of our system because it commutes with the stationary density matrix Eq.~(\ref{sigmastat}) calculated using only the tunnelling component of the Liouvillean. For a generic description of the Coulomb interaction on the molecule, corrections to $H_{\rm eff}$ given by the $8$ and $6$ particle excited states should be taken into account and the form of $H_{\rm eff}$ is modified. For the sake of simplicity we restrict here to the constant interaction model.  More details on the derivation and the discussion on the most general case are given instead in the Appendix~\ref{APP_omega_L}.

\begin{figure}[h]\centering
\includegraphics[width=0.3\textwidth]{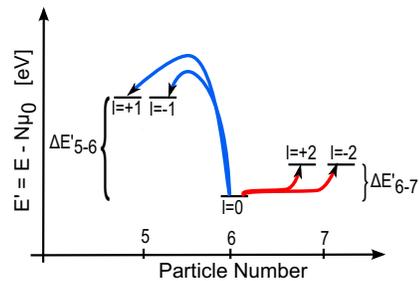}
\caption{(Color online) Together with a change in the energy, the transition from the 6-particle ground state to the 7-particle (5-particle) ground states is also associated with a change in the angular momentum of $\Delta l = \pm 2$ ($\Delta l = \pm 1$).}
\label{tunevent}
\end{figure}

\section{I-V characteristics and current maps of a benzene molecule}

\begin{figure}[h]
\includegraphics[width=0.5\textwidth]{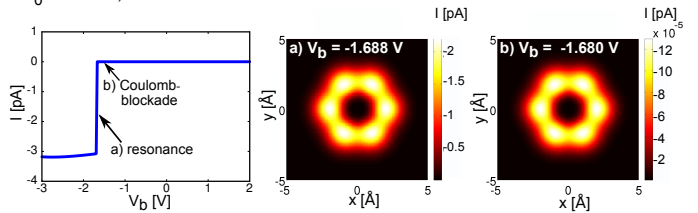}
\caption{(Color online) Current voltage characteristics and current maps associated to the neutral-anion transition. The current maps are calculated with $z_{\rm tip} - d = 5$\AA. Notice that the map in the Coulomb blockade region is just a rescaling of the one at resonance.}
\label{67_resonant}
\end{figure}

\begin{figure}[h!]\centering
\includegraphics[width=0.5\textwidth]{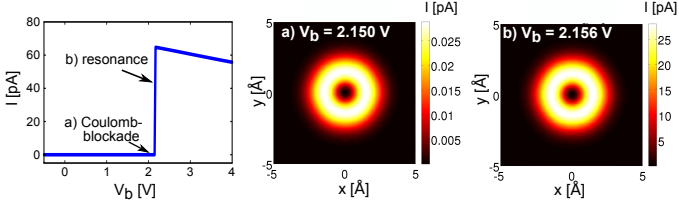}
\caption{(Color online) Current voltage characteristics and current maps associated to the neutral-cation transition. The current maps are calculated with $z_{\rm tip} - d = 5$\AA. Notice that the value of the current at resonance is much higher than the one relative to the neutral-anion case (see Fig.~\ref{67_resonant}). }
\label{65_resonant}
\end{figure}

\begin{figure}[h!]
\includegraphics[width=0.5\textwidth]{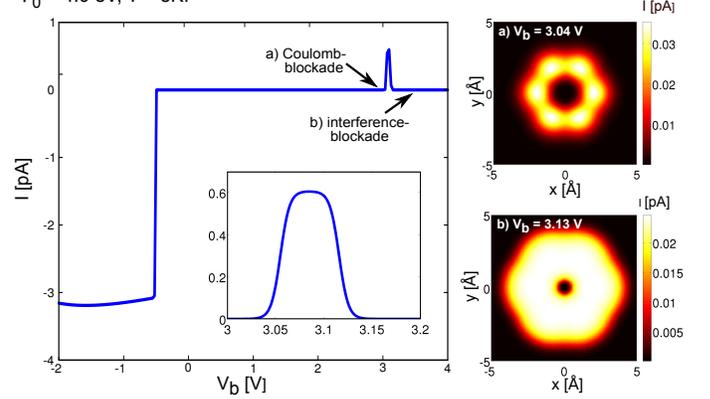}
\caption{(Color online) Current voltage characteristics and current maps associated to the neutral-anion transition. Interestingly the current map in the interference blockade region shows novel topographic features if compared with other maps involving the same states (see also Fig.~\ref{67_resonant}). \hlA{In the inset a zoom on the interference current peak is presented.}}
\label{67 interference}
\end{figure}

In the following discussion of the current voltage characteristics and current maps we only consider the ground state transition $|6_g 0 0\rangle \leftrightarrow | 7_g l \tau \rangle $ or $| 6_g 0 0\rangle \leftrightarrow | 5_g l \tau \rangle$. In Fig.~\ref{tunevent} we represent the corresponding energy levels as a function of the particle number for a particular choice of the work function (we assume $\Phi_0^T = \Phi_0^S$ so that the chemical potentials are the same at $V_b = 0$). In the tunnelling event the molecule changes its particle number, angular momentum and energy (see Fig.~\ref{tunevent}). All these changes leave their fingerprints in the current voltage characteristics and current maps presented in Figs.~\ref{67_resonant}-\ref{67 interference}.

In particular, the current is exponentially suppressed at small biases  (the so called ``in gap region'' of transport) due to the Coulomb blockade \cite{Beenakker_prb_91}. The bias at which current starts to flow corresponds to a resonant condition between the chemical potential in the source (or drain) lead and the difference in the energy $\Delta E$ between the many-body states participating to the transport. For this reason the current voltage characteristics (and the associated differential conductance traces) recorded with an STM junction represent a valuable spectroscopic tool to investigate the many-body spectrum of the molecule. One has to keep in mind nevertheless that i) the resonant bias depends on the value of the work function of the leads, ii) the bias drops very asymmetrically at the tip and substrate interface with an associated very different amount of energy available to the molecular transition. The shift in the position of the resonance with the work function can be observed by comparing the positions of the step in the current at negative biases in Fig.~\ref{67_resonant} and \ref{67 interference}.

In Fig.~\ref{67 interference} one can also observe how the same molecular transition (between the neutral and anionic molecule) gives signals at different biases if triggered by a substrate ($V_b > 0$) or a tip ($V_b < 0$) tunnelling event. A larger bias (in absolute value) is needed for a substrate transition since most of the bias drop concentrates at the tip interface. Moreover the current signal obtained at positive bias is a peak instead of a step due to an interference blocking effect analogous to the one discussed in [\onlinecite{Darau}]. In the interference blocking region the system is blocked into a particular linear combination of the $7$ particles ground states that can be populated from the substrate but cannot be depopulated towards the tip.

\begin{figure}[h!]
\includegraphics[width=0.5\textwidth]{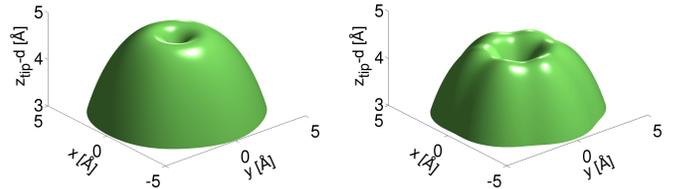}
\caption{(Color online) Constant current topographic images. The left panel refers to the neutral-cation resonance,  ($\phi_0 = 7 eV$, $V_b = 2.156 V$, $I = 300 pA$ ), the right panel, instead, to the neutral-anion resonance ($\phi_0 = 5 eV$, $V_b = -1.688 V$, $I = 100 pA$).}
\label{fig:Constant_I}
\end{figure}

The angular momentum channel involved in the transport depends on the difference in the angular momentum of the many-body states participating to the tunnelling events. The neutral-anion and neutral-cation transitions correspond to $\Delta l = \pm 2$ and $\Delta l = \pm 1$ respectively, cf. Fig.~\ref{tunevent}, thus involving different angular momentum channels. Since the lower is the angular momentum of the channel the larger are the rates, the current associated to the neutral-cation transition is larger than the one of the neutral-anion one, as it can be seen by comparing the resonant currents of Figs.~\ref{67_resonant} and \ref{65_resonant}. By comparing the same figures one finds also qualitative differences in  the constant heights current maps: yet another fingerprint of the different states involved in the transitions. The same differences are also confirmed by the constant current images presented in Fig.~\ref{fig:Constant_I}.

Finally, the current maps presented in Fig.~\ref{67 interference} suggests that also the interference effects have a topographic signature. The current map taken in the Coulomb blockade region is in fact qualitatively different from the one taken in the interference blockade.

\hlA{To conclude, a comparison with the widely applied Tersoff and Hamann (TH) theory \cite{Tersoff,Tersoff2,Tersoff_rev} is compulsory. In particular, for what concerns the current maps presented in Fig.~\ref{67_resonant} and Fig.~\ref{65_resonant}, we do not expect qualitative differences between the effectively single particle TH theory and our many-body approach. Yet, this is almost accidental for the following reasons: i) we decided for simplicity to describe the system using a constant interaction model in which the many-body states are single Slater determinants; ii) the initial and final many-body states of the tunneling event (e.g the neutral and anion ground states) fix the corresponding variation of angular momentum ($\Delta l = \pm 2$).  Consequently, in the particular case of benzene, only one single particle orbital contributes to the current formula given in Eq.~\eqref{subcurr}. In general, though, many Slater determinants are necessary to identify a single many-body state and many molecular orbitals would contribute to the transport. Moreover TH would not be able to address the interference blocking regime and the associated current maps since it is effectively a non interacting single particle theory.}

\section{Conclusions}

In this paper we presented an STM transport theory sufficiently general to be applied to any device consisting of a $\pi$-conjugated molecule weakly coupled both to the substrate and the tip. While the weak tunnelling coupling to the tip is a natural assumption in STM experiments, the weak coupling to the substrate is motivated by recent STM set-ups with substrates covered by a thin insulating film \cite{Repp, Repp2, Meyer_prl_2011}.

The essentially different geometry of the STM tip and the substrate is reflected in the respective tunnelling amplitudes, whose energy dependence induces, within a density matrix approach, characteristic non-constant tunnelling rate matrices. The latter play a central role in the Liouville operator, which determines the dynamics of the system, and in the current operator.

Interestingly, for these system, due to the different penetration lengths of the metallic states of the tip/substrate and the molecular orbitals into the corresponding tunnelling barriers, the tunnelling amplitudes cannot be calculated using the standard Tersoff and Hamann approach and an alternative method is proposed.

As an application of our general results we used a benzene molecule that enabled us to express the theory in the basis of the angular momentum $l$. The explicit calculation of the tunnelling rate matrices in the momentum basis shows a fundamental difference between the tip and substrate tunnelling dynamics. The delocalized tunnelling at the substrate happens via angular momentum channels (diagonal tunnelling matrices) while the localized tip mixes the angular momenta (off diagonal matrices).

A direct consequence of this different tunnelling scenario for the two leads is found in the current voltage characteristics. At voltages sufficiently large to lift the Coulomb blockade, interference blocking occurs when degenerate states participate to the transport. While the presence of degenerate states is a necessary condition for the interference, only the tip tunnelling can detect it due to its localized nature which mixes the angular momenta in the tunnelling event.

Moreover, also the STM surface-images can be calculated within our theory. By varying the work function of the substrate we show simulations of STM constant height current maps and constant current topographic images in which the transport is dominated either by neutral-anion or neutral-cation transitions. In particular, striking is the difference in the current maps obtained in the resonant and interference blocking regime although the same many body states participate to the transport (see Figs. \ref{67_resonant} and \ref{67 interference}).

\section*{Acknowledgements}
We thank prof.~Jascha Repp for the fruitful discussions. Moreover, we acknowledge financial support by the DFG within the research programs SPP 1243 and SFB 689.

\begin{appendix}

\section{Calculation of the overlap functions}\label{APPoverlap}

To calculate the tunnelling amplitude in equation (\ref{amplitude}) we need to calculate the overlap between the metal's wave function and the $p_z$-orbital. The latter is given, in the Gaussian description by\cite{Hehre, Jensen, NWchem}
\be\label{GaussOrb}
\langle \vec{r}|\alpha_G\rangle= n_G\sum_j \beta_j\, (\vec{r}-\vec{R}_{\alpha})\cdot \hat{e}_z \,\e^{-\alpha_j |\vec{r}-\vec{R}_\alpha|^2}\;,
\ee
where $n_G$ is the normalization factor which ensures $\int {\d}\vec{r}|\langle \vec{r}|\alpha\rangle|^2 = 1$, $\vec{R}_{\alpha}$ is the position of the atom $\alpha$ and $\hat{e}_z$ is the versor in the direction perpendicular to the plane of the molecule. Since the overlap is calculated as a function of the quantum number $k$ defining the lead wave function, we will call the bracket $\langle \chi k \sigma | \alpha \sigma \rangle$ overlap function.

\begin{figure}[h!]\centering
\includegraphics[width=0.4\textwidth]{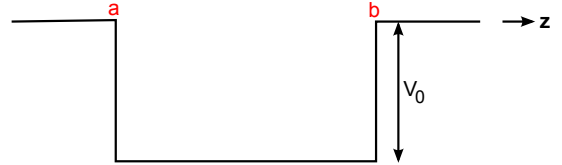}
\caption{Scheme of a 1-dimensional, finite potential well with borders $a$ and $b$ and depth $V_0$.}
\label{well_arbit}
\end{figure}

In our model both the tip and the substrate are described in the $z$ direction as potential wells \cite{Potential}. For future reference we report here the general expression for the eigenfunction of an arbitrary 1-dimensional potential well of depth $V_0$ and whose borders are $a$ and $b$, see Fig.~\ref{well_arbit}:

\begin{widetext}
\be\label{psiZ}
\Psi_{k_z}(z;a,b,V_0)=n_z\begin{cases}
            \e^{-\kappa a} \left[U\sin(k_za)+\cos(k_za)\right]\e^{\kappa z}\;,\quad &\text{if}\quad -\infty<z<a\\
            U\sin(k_zz)+\cos(k_zz)\;,\quad &\text{if}\quad a<z<b\\
            \e^{+\kappa b} \left[U\sin(k_zb)+\cos(k_zb)\right]\e^{-\kappa z}\;,\quad &\text{if}\quad b<z<\infty\;,\\
           \end{cases}
\ee
\end{widetext}
where $n_z$ ensures the normalization $\int {\rm d}z |\Psi_{k_z}(z)|^2 = 1$ and
\begin{equation*}
 U = \frac{k_z\sin(k_zb) - \kappa\cos(k_zb)}
          {k_z\cos(k_zb) + \kappa\sin(k_zb)}\;.
\end{equation*}
The occurring wave number reads $k_z=\sqrt{\frac{2m}{\hbar^2}\epsilon_z}$ and $\kappa = \sqrt{\frac{2m}{\hbar^2}V_0 - k_z^2}$, respectively. Due to the large size of the potential well compared to the Fermi wavelength we neglect the quantization of $k_z$ obtained by the corresponding eigenvalue equation.

We conclude this introductory part with the explicit calculation of an integral common to both the tip and substrate overlap functions. The integral is:

\begin{equation}
\label{eq:integral}
F_{k_z}(a,b,V_0,\alpha_j)= \int_{-\infty}^{+\infty}z\Psi_{k_z}(z+d;a,b,V_0)\e^{-\alpha_jz^2},
\end{equation}
where for simplicity we have omitted in $F$ the dependence on the parameter $d$. The integration yields:

\begin{widetext}
\be\label{z_overlap}
\begin{aligned}
&F_{k_z}(a,b,V_0,\alpha_j)
= \frac{n_z}{4\alpha_j^{\frac{3}{2}}}\\
&\times\Biggl\{\e^{-\frac{k_z^2}{4\alpha_j}} 2 {\rm Re}
	\Biggl[\e^{-\i k_z d}\left(1+\i U\right)\Biggl[\sqrt{\alpha_j}\left(\e^{-\alpha_j\left(a-d+\frac{\i k_z}{2\alpha_j}\right)^2} -
									    \e^{-\alpha_j\left(b-d+\frac{\i k_z}{2\alpha_j}\right)^2}\right)\\
                                               &\hspace{4.7cm}- \frac{\i k_z \sqrt{\pi}}{2}\left({\rm erf}\left[\sqrt{\alpha_j}\left(b-d+\frac{\i k_z}{2\alpha_j}\right)\right]-
				     {\rm erf}\left[\sqrt{\alpha_j}\left(a-d+\frac{\i k_z}{2\alpha_j}\right)\right] \right)\Biggr]\Biggr]\\
&\qquad\;+\e^{\frac{\kappa^2}{4\alpha_j}} \Biggl[ -A\e^{+\kappa d}\left(2\sqrt{\alpha_j}\e^{-\alpha_j\left(a-d-\frac{\kappa}{2\alpha_j}\right)^2} -
					 \kappa\sqrt{\pi}\left(1+ {\rm erf}\left[\sqrt{\alpha_j}\left(a-d-\frac{\kappa}{2\alpha_j}\right)\right]\right)\right)\\
&\hspace{1.9cm}\,+B \e^{-\kappa d}\left(2\sqrt{\alpha_j}\e^{-\alpha_j\left(b-d+\frac{\kappa}{2\alpha_j}\right)^2} -
					 \kappa\sqrt{\pi}\left(1- {\rm erf}\left[\sqrt{\alpha_j}\left(b-d+\frac{\kappa}{2\alpha_j}\right)\right]\right)\right)\Biggr]\Biggr\},
\end{aligned}
\ee
\end{widetext}
where we used the abbreviations
\begin{gather*}
A = \e^{-\kappa a}\left[U\sin(k_za)+\cos(k_za)\right]\;,\\
B = \e^{+\kappa b}\left[U\sin(k_zb)+\cos(k_zb)\right]\;.
\end{gather*}
In  equation (\ref{z_overlap}) the error function ${\rm erf}[\zeta]$ with $\zeta \in \mathbb{C}$ arises several times. It is defined as the integral of the normal distribution from $0$ to $\zeta$ scaled
such that ${\rm erf}[\pm \infty]=\pm 1$:
\begin{equation*}
 {\rm erf}[\zeta]=\frac{2}{\sqrt{\pi}}\int_0^\zeta\e^{-t^2}\d t\;
\end{equation*}
and it is an entire function valid for real- and complex valued numbers \cite{Abramowitz}.
Furthermore there holds
\begin{equation*}
\frac{2}{\sqrt{\pi}}\int_{\zeta_1}^{\zeta_2}\e^{-t^2}\d t = {\rm erf}[\zeta_2] -{\rm erf}[\zeta_1]\;.
\end{equation*}

Both  for the wave function $\Psi_{k_z}$ and the integral $F_{k_z}$ the tip and the substrate cases are obtained by the substitutions (see also the triple-well in Fig.~\ref{STMandWELL})

\begin{equation*}
 {\rm sub}:
 \begin{cases}
 a\rightarrow z_0\\
 b\rightarrow 0\\
 V_0 \rightarrow -\varepsilon_0^S
 \end{cases}
 \quad
 {\rm tip}:
 \begin{cases}
 a\rightarrow z_{\rm tip}\\
 b\rightarrow z_{\rm end}\\
 V_0 \rightarrow -\varepsilon_0^T
 \end{cases}
\end{equation*}

\subsection{Overlap molecule-substrate}

Let us consider the substrate case in which, for the sake of simplicity, we neglect in the following the spinor component of the substrate and atomic states. According to the model given in the main text and sketched in Fig.~\ref{STMandWELL}, the substrate's wave function is given by
\begin{equation}\label{wfsub}
\langle x,y,z|S \vec{k} \rangle=\frac{1}{\sqrt{S}}
\e ^{+\i (k_x x + k_y y)}\Psi_{k_z}(z;z_0,0,-\varepsilon_0^S)\;,
\end{equation}
where $k_{x/y/z}=\sqrt{\frac{2m}{\hbar^2}\varepsilon_{x/y/z}}$ and $S$ is the area of the surface of the substrate on which the molecule lies.
The exponentials in Eq.~\eqref{wfsub} stem from using no confinement to describe the substrate in the $x$ and $y$ direction and periodic boundary conditions. Due to the large size of the substrate in all the three directions if compared with the Fermi wavelength $\lambda_{F} = \sqrt{\hbar^2/(2m\varepsilon_F)}$ we neglect the momentum quantization in all three directions. By setting the origin of the coordinate system in $\vec{R}_{\alpha}$ and performing the Gaussian integrals in the $x$ and $y$ direction one easily obtains:

\be\begin{aligned}\label{FourierG}
\langle S \vec{k}|\alpha_G\rangle = &
\e^{-\i\vec{k}_{||}\cdot\vec{R}_\alpha}
\frac{n_G}{\sqrt{S}}
\sum_j\frac{\pi\beta_j}{\alpha_j}\e^{-\frac{k_{||}^2}{4\alpha_j}}\\
&\times\int_{-\infty}^{+\infty} \!\!\! {\rm d}z\, z\, \Psi_{k_z}(z+d,z_0,0,-\varepsilon_0^T)\e^{-\alpha_j z^2}\\
= & \e^{-\i\vec{k}_{||}\cdot\vec{R}_\alpha}
\frac{n_G}{\sqrt{S}}
\sum_j\frac{\pi\beta_j}{\alpha_j}\e^{-\frac{k_{||}^2}{4\alpha_j}}\\
&\times F_{k_z}(z_0,0,-\varepsilon_0^S,\alpha_j):=\e^{-\i\vec{k}_{\parallel}\cdot\vec{R}_\alpha}O_S(\vec{k}),
\end{aligned}
\ee
where $\vec{k}_{||}\cdot\vec{R}_\alpha = k_x x_\alpha + k_y y_\alpha$ and the integral in the $z$ direction has been performed with the help of Eqs.~\eqref{eq:integral} and \eqref{z_overlap}. Notice the suppression of the overlap for high values of the parallel component of the momentum $|\vec{k}_{\parallel}|$ in the substrate wave function given by the gaussian prefactor and also the phase factor which depends on the position of the carbon atom $\vec{R}_{\alpha}$ and on $\vec{k}_{\parallel}$.

\begin{figure}\centering
\includegraphics[width=0.4\textwidth]{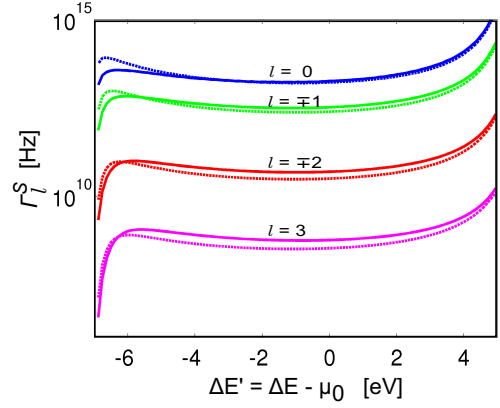}
\caption{(Color online) Tunneling-rates obtained by using Slater-type orbitals (solid lines) and Gaussian orbitals (dashed lines). The rates are calculated for a substrate-molecule distance $d = 3 \mathring{A}$. In the Slater-type orbital $Z_{\rm eff} = 2$.}
\label{rateCompare}
\end{figure}

Instead of using a Gaussian $p_z$ orbital we can also use a Slater-type orbital \cite{Slater, Belkic}:
\be\begin{aligned}
\langle \vec{r}|\alpha_{S}\rangle=&
\frac{1}{2{\sqrt{6}}}
\left(\frac{Z_{\rm eff}}{a_0}\right)^\frac{5}{2}\,
(\vec{r}- \vec{R}_\alpha) \cdot \hat{e}_z
\e^{-\frac{Z_{eff}}{a_0} |\vec{r}-\vec{R}_\alpha|},
\end{aligned}\ee
where $a_0 = 0.53 \mathring{A}$ is the Bohr radius and $Z_{\rm eff}$ is a fitting parameters that takes into account the screening of the nuclear potential given by the core electrons. In Fig.~\ref{rateCompare} we show the substrate-tunnelling rates for the different benzene molecular orbitals calculated according to  Eq.~\eqref{Trate}. We compare the rates obtained using Gaussian and Slater-type orbitals using a distance $d = 3 \mathring{A}$ between the end of the metallic well (the substrate) and the plane of the molecule. As one can see the two results are in good agreement. The discrepancy between the two descriptions depends nevertheless on the distance $d$ due to the difference in the tails of the Slater and Gaussian descriptions of the $p_z$ orbital. A good agreement is reached in the range of $d$ we are interested in ($d = 1\mathring{A} -6 \mathring{A}$).

\subsection{Overlap molecule-tip}
We continue with the calculation of the tip-orbital overlap. The atomic wave function is described again by the Gaussian orbitals given in Eq.~(\ref{GaussOrb}). The tip is modeled assuming a harmonic confinement in $x$ and $y$ direction, and a quantum well for the $z$ one. The overlap reads:
\be
\begin{split}
\langle x,y,z | T k_z \rangle= \sqrt{\frac{m\omega}{\pi\hbar}}\Psi_{k_z}(z;z_{\rm tip},z_{\rm end},-\varepsilon_0^T)\\
\times \e^{-\frac{m\omega}{2\hbar}\left((x-x_{\rm tip})^2+(y-y_{\rm tip})^2\right)}\;.
\end{split}
\ee
The overlap function is a three dimensional integral which, in Cartesian coordinates, reads:
\be\label{tipOV}
\begin{aligned}
&\langle Tk_z|\alpha_G\rangle = n_G \sqrt{\frac{m\omega}{\pi\hbar}}\sum_j\beta_j\\
&\quad
\int_{-\infty}^\infty \!\!\!\d x
\int_{-\infty}^\infty\!\!\! \d y
\int_{-\infty}^\infty\!\!\! \d z
\,\e^{-\alpha_j\left[(x-x_\alpha)^2+(y-y_\alpha)^2+(z-d)^2\right]}\\
&\qquad\qquad
\times(z-d)\Psi_{k_z}(z;z_{\rm tip},z_{\rm end},-\varepsilon_0^T)\\
&\qquad\qquad
\times\e^{-\frac{m\omega}{2\hbar}
\left[(x-x_{\rm tip})^2+(y-y_{\rm tip})^2\right]}\;,
\end{aligned}
\ee
where we have already set $z_{\alpha} = d$ $\forall \alpha$. We shift again the origin of the coordinates to the center of the $p_z$ orbital, $\vec{R}_\alpha$, and perform the gaussian integrals in the $x$ and $y$ direction. Moreover it is convenient to introduce  new variables describing the tip-atom distance
$ \Delta x = x_{\rm tip} - x_\alpha, \,
\Delta y = y_{\rm tip} - y_\alpha$. The resulting overlap function reads

\be \label{eq:O_T_final}
\begin{aligned}
 \langle Tk_z|\alpha_G\rangle =& n_G\sqrt{\frac{m\omega}{\pi\hbar}}\sum_j\frac{\beta_j\pi}{\alpha_j+\frac{m\omega}{2\hbar}}
\,\e^{-\frac{m\omega\alpha_j}{2\hbar\alpha_j+m\omega}
\left(\Delta x^2 + \Delta y ^2\right)}\\
&\times F_{k_z}(z_{\rm tip},z_{\rm end},-\varepsilon_0^T,\alpha_j)\\ &:=O_T(k_z,\vec{R}_{\rm tip}-\vec{R}_\alpha)
\end{aligned}
\ee
and concludes this section dedicated to the explicit calculation of the overlap functions.

\section{The stationary density matrix}\label{AppsigmaStat}

In Eq.~\eqref{sigmastat} we only gave the generic form of the stationary density matrix $\sigma_{\rm stat}$ for an orbitally degenerate subspace. In this section we will show how to calculate it and finally give the complete result for a specific example. For the sake of simplicity we concentrate on the transitions $6_g\leftrightarrow 7_g$, but the calculation can be easily reproduced for all other transitions. The reduced density matrix $\sigma_{\rm stat}$ for the specific subspace that we are considering is composed of a single-element sub-block associated to the $6$ particle ground state, a $2\times2$ sub-block associated to the subspace $\{|7_g +2\uparrow\rangle, \, |7_g -2\uparrow\rangle\}$ and finally a $2\times2$ sub-block  relative to the space ${\rm span}\{|7_g +2 \downarrow\rangle, \, |7_g -2 \downarrow\rangle\}$. Since we are interested in orbital (but not spin) coherences the Liouvillean is a linear operator of dimension $9\times9$. We choose the basis:
\begin{equation}
|6_g \rangle\rangle,\quad\\
\begin{cases}
|7_g \uparrow; +2,+2 \rangle\rangle\\
|7_g \uparrow; -2,-2 \rangle\rangle\\
|7_g \uparrow; +2,-2 \rangle\rangle\\
|7_g \uparrow; -2,+2 \rangle\rangle\\
\end{cases}\!\!\!,
\begin{cases}
|7_g \downarrow; +2,+2\rangle\rangle\\
|7_g \downarrow; -2,-2\rangle\rangle\\
|7_g \downarrow; +2,-2\rangle\rangle\\
|7_g \downarrow; -2,+2\rangle\rangle
\end{cases}
\end{equation}
where the notation $|\phantom{m}\rangle\rangle$ denotes a vector in the density matrix space. We organize the tunnelling Liouvillean in the following form:

\begin{equation}
\label{eq:Liouville}
(\mathcal{L}_{\rm tun})_{6_g7_g} = \left(
                                     \begin{array}{ccc}
                                       \mathcal{L}_{66} & \mathcal{L}_{67\uparrow} & \mathcal{L}_{67\downarrow} \\
                                       \mathcal{L}_{7\uparrow6} & \mathcal{L}_{7\uparrow7\uparrow} & 0 \\
                                       \mathcal{L}_{7\downarrow6} & 0 & \mathcal{L}_{7\downarrow7\downarrow} \\
                                     \end{array}
                                   \right),
\end{equation}
where $\mathcal{L}_{66} = -4\left(f_T^+\Gamma^T + f_S^+\Gamma^S\right)$ is the depopulation rate of the 6 particle ground state and the coefficients $\Gamma^{S/T}$ stand for the diagonal elements of the tunnelling rates of the substrate or the tip. Moreover, the rates and the Fermi functions are calculated at the same energy $\delta E = E_{7_g} - E_{6_g}$. The other elements of the matrix $(\mathcal{L}_{\rm tun})_{6_g7_g}$ are matrices themselves. In particular:

\begin{equation}
\begin{split}
\mathcal{L}_{67\uparrow} =  \mathcal{L}_{67\downarrow} =&
f_S^-\Gamma^S \left(
                             \begin{array}{cccc}
                               1 & 1 &  0 & 0\\
                             \end{array}
              \right)\\
&+
f_T^-\Gamma^T
\left(
                             \begin{array}{cccc}
                               1 &
                               1 &
                    \e^{+2\i\phi_2} &
                    \e^{-2\i\phi_2} \\
                             \end{array}
                           \right)
\end{split}
\end{equation}
are the population ``rates'' of the 6 particle ground state starting from the states $|7_g l\uparrow\rangle $ and $|7_g l\downarrow\rangle$, while

\begin{equation}
\begin{split}
\mathcal{L}_{7\uparrow6} = \mathcal{L}_{7\downarrow6} = &
f_S^+\Gamma^S
                            \left(
                             \begin{array}{cccc}
                               1 & 1 & 0 & 0 \\
                             \end{array}
                           \right)^T\\
                           &+
 f_T^+\Gamma^T
                           \left(
                             \begin{array}{cccc}
                             1 &
                             1  &
                    \e^{-2\i\phi_2} &
                    \e^{+2\i\phi_2} \\
                             \end{array}
                           \right)^T
\end{split}
\end{equation}
are the population ``rates'' of the states $|7_g l\uparrow\rangle$ and $|7_g l\downarrow\rangle$ starting from the state $|6_g\rangle$. Finally
\begin{widetext}
\begin{equation}
\mathcal{L}_{7\uparrow7\uparrow} = \mathcal{L}_{7\downarrow7\downarrow} =
-f_T^-\Gamma^T
\left(
      \begin{array}{cccc}
      1 & 0 & \e^{+2\i\phi_2}/2 & \e^{-2\i\phi_2}/2 \\
      0 & 1 & \e^{+2\i\phi_2}/2 & \e^{-2\i\phi_2}/2 \\
      \e^{-2\i\phi_2}/2 & \e^{-2\i\phi_2}/2 & 1 & 0 \\
      \e^{+2\i\phi_2}/2 & \e^{+2\i\phi_2}/2 & 0 & 1 \\
      \end{array}
\right)
-f_S^-\Gamma^S
\left(
      \begin{array}{cccc}
      1 & 0 & 0 & 0 \\
      0 & 1 & 0 & 0 \\
      0 & 0 & 1 & 0 \\
      0 & 0 & 0 & 1 \\
      \end{array}
\right)
\end{equation}
\end{widetext}
is the depopulation ``rate'' of the states $|7_g l\uparrow\rangle$ and $7_g l \downarrow\rangle$ towards the 6 particle ground state.

The stationary solution of the Generalized Master Equation Eq.~\eqref{GME} is found by calculating the null space of the Liouville operator. Here we restrict ourselves to the operator $\mathcal{L}_{\rm tun}$ describing the sequential tunnelling dynamics.    A discussion about the relevance of the commutator with the effective Hamiltonian is left to the last appendix.
If the leads are not superconductors, non magnetic or with parallel polarization and weakly coupled to the molecule, the stationary density matrix is block diagonal in particle number, energy and spin. Thus, the stationary solution which corresponds to the Liouvillean given in Eq.~\eqref{eq:Liouville} can be cast into the form:

\begin{equation}
\sigma_{\rm stat} = \left(
                      \begin{array}{ccc}
                        \sigma_{6_g} & 0 & 0 \\
                        0 & \sigma_{7_g \uparrow} & 0 \\
                        0 & 0 & \sigma_{7_g\downarrow} \\
                      \end{array}
                    \right)
\end{equation}
where the 7 particle subblocks, when written  in the basis $\{|7_g +\!2\,\tau \rangle, |7_g -\!2\,\tau \rangle \}$, read:

\begin{equation}
\begin{split}
\sigma_{7_g\uparrow} &= \sigma_{7_g\downarrow} = \left(
                                                  \begin{array}{cc}
                                                    A & B\e^{-2\i\phi_2} \\
                                                    B\e^{+2\i\phi_2} & A \\
                                                  \end{array}
                                                \right)\\
\end{split}
\end{equation}
with
\begin{equation}
\begin{split}
\sigma_{6_g} &= \frac{f^-_S \Gamma^S (f^-_S \Gamma^S + 2 f^-_T \Gamma^T)}{N}\;,\\
A &= \frac{f^-_T \Gamma^T f^+_S \Gamma^S + f^-_S \Gamma^S (f^+_S \Gamma^S + f^+_T \Gamma^T)}{N}\;,\\
B &=\frac{\Gamma^S\Gamma^T(f^-_S  f^+_T - f^-_T f^+_S)}{N}\;,\\
\end{split}
\end{equation}
and  the normalization $N$ defined by the relation ${\rm Tr}\sigma_{\rm stat} = 1$. This result is worth some further analysis. First of all it is interesting to notice that $B = 0$ only if at least one of the following conditions is satisfied i) $\Gamma^S = 0$ which is never happening, ii) $\Gamma^T = 0$ which holds if $\vec{R}_{\rm tip}$ is on the main rotational axis of benzene, iii) $f^+_T/f^-_T = f^+_S/f^-_S$ which is satisfied only in equilibrium when $\mu_T = \mu_S$. This analysis shows how the interference between states with different angular momenta is ubiquitous in the molecular junction. Eventually, it is easy to prove that the eigenvalues of the stationary density matrix are $\sigma_{6_g}$, $A+B$ and $A-B$. The ratio between these eigenvalues gives a key to the physical interpretation of the stationary density matrix. In fact:

\begin{equation}
\label{eq:Pop_ratios}
\begin{split}
\frac{A+B}{\sigma_{6_g}} &=
\frac{\Gamma^Sf^+_S + 2\Gamma^Tf^+_T}
{\Gamma^Sf^-_S + 2\Gamma^Tf^-_T}\; ,\\
\frac{A-B}{\sigma_{6_g}} &= \frac{f^+_S}{f^-_S} = \e^{-\beta(\Delta E - \mu_S)}\; ,
\end{split}
\end{equation}
which can be interpreted as follows: $\sigma_{7g}^D :=A-B$ is the occupation of the 7 particle state $|{7_g} D\tau \rangle$ which is decoupled from the tip and coupled to the 6 particle ground states only via tunnelling events happening at the molecule substrate interface. For this reason the ratio $\sigma_{7g}^D/\sigma_{6_g}$ is the same as the one obtained in thermal equilibrium with the substrate. On the other hand $\sigma_{7g}^C :=A+B$ is the population of the 7 particle state $|{7_g} C\tau \rangle$ which can exchange particles both at the molecule-substrate and at the molecule-tip interfaces. In particular, the rate of exchange for the state $|7g C \tau\rangle$ is double than the rate of exchange of the angular momentum states $|7g l \tau\rangle$ (see Eq.~\eqref{tipmatrix}). The detailed balance gives immediately the first relation in Eq.~\eqref{eq:Pop_ratios}.

\section{Phase of the tunnelling amplitude}\label{app:phase}

The phase of the tunnelling amplitude between a benzene molecular orbital and a tip state plays an important role in the calculation of the transport characteristics of the STM junction. In this section we derive the approximate formula describing this phase given by Eq.~\eqref{eq:Phase_approx}, and also its limit of validity.
Due to the cylindrical symmetry of the tip wave function, for the overlap function with the atomic wave function $O_T(k_z,\vec{R}_{\rm tip}-\vec{R}_{\alpha})$ it holds:
\begin{equation}
\label{eq:OVT}
O_T(k_z,\vec{R}_{\rm tip}-\vec{R}_{\alpha}) =
f(k_z,z_{\rm tip},|\vec{r}_{\rm tip} - \vec{r}_{\alpha}|),
\end{equation}
where $f$ is a real function (see Eq.~\eqref{eq:O_T_final} in appendix A) and we have introduced cylindrical coordinates with the origin in the center of the molecule and the $z$ axis perpendicular to the molecular plane. Every point $\vec{R}$ in the space is thus described by the triplet $(z,r,\theta)$ and we fix $\theta = 0$ along the radius intersecting the atom with $\alpha = 0$ (see Fig.~\ref{SymmetryPhase}). Finally, we have defined  $\vec{r}$ to be the projection of $\vec{R}$ in the plane of the molecule. It follows immediately that

\be
|\vec{r}_{\rm tip} - \vec{r}_{\alpha}| = \sqrt{a^2 + r_{\rm tip}^2 - 2ar_{\rm tip} \cos(\theta_{\rm tip}-\theta_\alpha)},
\end{equation}
where $a$ is the distance between the carbon atoms and the center of the molecule, and $\theta_\alpha = (2\pi/6)\alpha $ with $\alpha = 0,\ldots,5$. Combining Eqs. \eqref{eq:OVT}, \eqref{eq:ampliTIP}, \eqref{eq:basissub} and \eqref{eq:phasetip} we obtain:

\begin{equation}
\phi_l(\vec{R}_{\rm tip})
= {\rm arg} \left\{ \sum_{\alpha}
f[z_{\rm tip},r_{\rm tip},\cos(\theta_\alpha-\theta_{\rm tip})]\e^{\i l \theta_\alpha}\right\}
\end{equation}
and, consequently:

\begin{equation}
\phi_l(\vec{R}_{\rm tip})- l\theta_{\rm tip} = {\rm arg} \left\{ \sum_{\alpha}
f(z_{\rm tip},r_{\rm tip},\cos\phi_\alpha)\e^{\i l \phi_\alpha}\right\},
\end{equation}
where $\phi_\alpha = \theta_{\alpha}-\theta_{\rm tip}$. If now we expand $f$ in the Taylor series:

\begin{equation}
f(z_{\rm tip},r_{\rm tip},\cos \phi_\alpha) = \sum_{n = 0}^{\infty} \left.\frac{f^{(n)}}{n!}\right|_{(z_{\rm tip},r_{\rm tip},0)} (\cos \phi_\alpha)^n
\end{equation}
we reduce the problem to the evaluation of the functions

\begin{equation}
g_{nl}(\theta_{\rm tip}) = \sum_{\alpha}
[\cos(\theta_\alpha - \theta_{\rm tip})]^n\e^{\i l(\theta_\alpha - \theta_{\rm tip})},
\end{equation}
which is easily done by means of the Euler formula for the cosine and the binomial theorem. The solution reads:

\begin{equation}
\label{eq:g_nl}
\begin{split}
g_{nl}(\theta_{\rm tip}) =& \frac{6}{2^n}\sum_{c \in \mathbb{Z}}
\left(
\begin{array}{c}
  n \\
  \frac{n+6a-l}{2}
\end{array}
\right)
\e^{-\i 6c\theta_{\rm tip}} \\
& \times \left|\cos\left[\frac{\pi}{2}(n+6c-l)\right]\right|\\
& \times \theta(n+6c-l+2)\theta(n-6c+l+2),
\end{split}
\end{equation}
with $\theta(x) = 1$ if $x > 0$ and 0 elsewhere. By analyzing Eq.~\eqref{eq:g_nl} we obtain the following general properties:
i) If $\theta_{\rm tip} = n\pi/6$, with $n \in \mathbb{N}$, the function $g_{nl}(\theta_{\rm tip})$ is real, thus Eq.~\eqref{eq:Phase_approx} is exact when $\vec{R}_{\rm tip}$ is on the planes perpendicular to the molecule passing through the center of the molecule and one of the atoms or the center and the middle point of a carbon-carbon bond. ii) $g_{n1} = 0$ if $n$ is even and $g_{n2} = 0$ if $n$ is odd, $\forall \theta_{\rm tip}$. iii) $g_{n1}$ is real for $n \leq 4$ and $g_{n2}$ is real for $n \leq 3$. The combination of the observation ii) and iii) supports the validity of  Eq.~\eqref{eq:Phase_approx} on the entire space.

\section{The effective Hamiltonian}\label{APP_omega_L}

In this section we analyze the form of the effective Hamiltonian $H_{\rm eff}$ introduced in Eq.~\eqref{eq:H_eff} both in the case of a constant interaction model or a more generic model for the interaction. The discussion will always be restricted to the subspace spanned by the 7 particle ground states in which the effective Hamiltonian reduces to a $4\times 4$ matrix whose generic element is $\langle 7_g l \tau | H_{\rm eff} | 7_g l' \tau' \rangle$.

Since $[H_{\rm m},S_z] = 0$ it follows immediately that $H_{\rm eff}$ is diagonal in the spin quantum number. Moreover one proves the following relations:

\begin{equation}
\label{eq:H_eff_sym}
\begin{split}
\langle 7_g l\tau | H_{\rm eff} | 7_g l' \tau \rangle &=
\langle 7_g l\bar{\tau} | H_{\rm eff} | 7_g l' \bar{\tau} \rangle,\\
\langle 7_g l\tau | H_{\rm eff} | 7_g l \tau \rangle &=
\langle 7_g \bar{l}\bar{\tau} | H_{\rm eff} | 7_g \bar{l} \bar{\tau} \rangle,\\
\end{split}
\end{equation}
which ensures i) that the two spin subblocks are identical and ii) that the diagonal elements in each of the two subblocks are equal.
In order to prove the relations given in Eq.~\eqref{eq:H_eff_sym}
it is useful to introduce the symmetry operations $U_{\rm spin}$ and $U_{\rm orb}$ defined as follows:

\begin{equation}
\begin{split}
d_{l\bar{\sigma}} &= U_{\rm spin} d_{l\sigma} U_{\rm spin}^\dagger,\\
d_{\bar{l}\sigma} &= U_{\rm orb} d_{l\sigma} U_{\rm orb}^\dagger.
\end{split}
\end{equation}

The proof of the first relation in \eqref{eq:H_eff_sym} is readily given:

\begin{widetext}
\begin{equation}
\begin{split}
\langle 7_g n \tau | H_{\rm eff} |7_g n' \tau \rangle
=&
\frac{1}{2\pi}\sum_{ll'\chi}
\left[
\langle 7_g n \tau |
d^\dagger_{l\sigma}
\Gamma^\chi_{ll'}(E_{7_g}-H_{\rm m})
p_\chi(E_{7_g}-H_{\rm m})
d_{l'\sigma}
|7_g n' \tau \rangle \right.\\
&\left.+
 \langle 7_g n \tau |
d_{l'\sigma}
\Gamma^\chi_{ll'}(H_{\rm m}-E_{7_g})
p_\chi(H_{\rm m}-E_{7_g})
d^\dagger_{l\sigma}
|7_g n' \tau \rangle
\right]\\
=&
\frac{1}{2\pi}\sum_{ll'\chi}
\left[
\langle 7_g n \bar{\tau} |
d^\dagger_{l\bar{\sigma}}
\Gamma^\chi_{ll'}(E_{7_g}-H_{\rm m})
p_\chi(E_{7_g}-H_{\rm m})
d_{l'\bar{\sigma}}
|7_g n' \bar{\tau} \rangle \right.\\
&\left.+
 \langle 7_g n \bar{\tau} |
d_{l'\bar{\sigma}}
\Gamma^\chi_{ll'}(H_{\rm m}-E_{7_g})
p_\chi(H_{\rm m}-E_{7_g})
d^\dagger_{l\bar{\sigma}}
|7_g n' \bar{\tau} \rangle
\right] = \langle 7_g n \bar{\tau} | H_{\rm eff} |7_g n' \bar{\tau} \rangle,
\end{split}
\end{equation}
\end{widetext}
where for the second equality we have introduced the identity operators $U^\dagger_{\rm spin}U_{\rm spin}$ before and after the operators $d_{l\sigma}$ and $d^\dagger_{l'\sigma}$. The last equality is obtained by replacing $\bar{\sigma} \to \sigma$ in the sum and remembering that $\Gamma^\chi_{ll'}$ is independent of the spin of the electron in the lead. The second relation in \eqref{eq:H_eff_sym} is obtained in an analogous way:

\begin{widetext}
\begin{equation}
\begin{split}
\langle 7_g n \tau | H_{\rm eff} |7_g n \tau \rangle
=&
\frac{1}{2\pi}\sum_{l\chi}
\left[
\langle 7_g n \tau |
d^\dagger_{l\sigma} \Gamma^\chi_{ll}(E_{7_g}-H_{\rm m})
p_\chi(E_{7_g}-H_{\rm m})d_{l\sigma}
|7_g n \tau \rangle \right.\\
&\left.+
 \langle 7_g n \tau |
d_{l\sigma} \Gamma^\chi_{ll}(H_{\rm m}-E_{7_g})
p_\chi(H_{\rm m}-E_{7_g})d^\dagger_{l\sigma}
|7_g n \tau \rangle
\right]\\
=&
\frac{1}{2\pi}\sum_{l\chi}
\left[
\langle 7_g \bar{n} \tau |
d^\dagger_{\bar{l}\sigma} \Gamma^\chi_{ll}(E_{7_g}-H_{\rm m})
p_\chi(E_{7_g}-H_{\rm m})d_{\bar{l}\sigma}
|7_g \bar{n} \tau \rangle \right.\\
&\left.+
 \langle 7_g \bar{n} \tau |
d_{\bar{l}\sigma} \Gamma^\chi_{ll}(H_{\rm m}-E_{7_g})
p_\chi(H_{\rm m}-E_{7_g})d^\dagger_{\bar{l}\sigma}
|7_g \bar{n}\tau \rangle
\right] = \langle 7_g \bar{n} \tau | H_{\rm eff}
|7_g \bar{n}\tau \rangle,
\end{split}
\end{equation}
\end{widetext}
where the first equality is obtained by removing the sum over $l'$ since the Hamiltonian $H_{\rm m}$ conserves the $z$ projection of the angular momentum, the second equality proceeds instead by inserting the identities $U^\dagger_{\rm orb}U_{\rm orb}$ before and after the operators $d_{l\sigma}$ and $d^\dagger_{l'\sigma}$. Finally, in the last equality, we have redefined $\bar{l} \to l$ and used the symmetry property of the rate matrices $\Gamma^\chi_{ll}= \Gamma^\chi_{\bar{l}\bar{l}}$.

For the analysis of the off diagonal elements of $H_{\rm eff}$ within a single spin subblock we have to distinguish between the substrate and the tip case. In the substrate case $\Gamma^{S}_{ll'} \propto \delta_{ll'}$ which directly implies that also the component of $H_{\rm eff}$ given by the coupling to the substrate is diagonal and, due to the second relation in \eqref{eq:H_eff_sym} proportional to the identity matrix and thus irrelevant for the dynamics of the molecule.

Thus, let us concentrate on the tip contribution. It is possible to  demonstrate that:

\begin{equation}
\label{eq:H_eff_OffDiag}
\langle 7_g \,+2\,\tau |H_{\rm eff}^T | 7_g -\!2\, \tau  \rangle =
A \e^{-2\i\phi_2} + B \e^{-\i\phi_1},
\end{equation}
where we have introduced the notation $H_{\rm eff}^T$ to indicate the component of $H_{\rm eff}$ with $\chi = T$, $\phi_1$ and $\phi_2$ are the phases of the tunnelling amplitudes calculated in the previous section. Finally, $A,\,B \in \mathbb{R}$ are given by

\begin{widetext}
\begin{equation}
\label{eq:AB}
\begin{split}
A =& \frac{1}{2\pi} \sum_\sigma
\left[
\langle 7_g \,2\,\tau |
d^\dagger_{2\sigma}
\left|\Gamma^T_{2,-2} (E_{7_g}-H_{\rm m})\right|
p_T(E_{7_g}-H_{\rm m})
d_{-2\sigma}
| 7_g -\!2\,\tau \rangle\right.\\
&\left.+
\langle 7_g \,2\,\tau |
d_{-2\sigma}
\left|\Gamma^T_{2,-2} (H_{\rm m}-E_{7_g})\right|
p_T(H_{\rm m}-E_{7_g})
d^\dagger_{2\sigma}
| 7_g -\!2\,\tau \rangle
\right]\;,\\
B =&  \frac{1}{\pi} {\rm Re}\sum_\sigma
\left[
\langle 7_g \,2\,\tau |
d^\dagger_{1\sigma}
\left|\Gamma^T_{1 3} (E_{7_g}-H_{\rm m})\right|
p_T(E_{7_g}-H_{\rm m})
d_{3\sigma}
| 7_g -\!2\,\tau \rangle\right.\\
&\left.+
\langle 7_g \,2\,\tau |
d_{3\sigma}
\left|\Gamma^T_{1 3} (H_{\rm m}-E_{7_g})\right|
p_T(H_{\rm m}-E_{7_g})
d^\dagger_{1\sigma}
| 7_g -\!2\,\tau \rangle
\right].
\end{split}
\end{equation}
\end{widetext}
The proof of Eq.~\eqref{eq:H_eff_OffDiag} proceeds as follows. Let us start from the definition of the off diagonal matrix element:

\begin{widetext}
\begin{equation}
\label{eq:H_eff_OffDiag_def}
\begin{split}
\langle 7_g \,2\,\tau | H_{\rm eff}^T |7_g -2 \tau \rangle =&
\frac{1}{2\pi}\sum_{ll'\chi}
\left[
\langle 7_g \,2\,\tau |
d^\dagger_{l\sigma}
\Gamma^\chi_{ll'}(E_{7_g}-H_{\rm m})
p_\chi(E_{7_g}-H_{\rm m})
d_{l'\sigma}
|7_g -2 \tau \rangle \right.\\
&\left.+
 \langle 7_g -2 \tau |
d_{l'\sigma}
\Gamma^\chi_{ll'}(H_{\rm m}-E_{7_g})
p_\chi(H_{\rm m}-E_{7_g})
d^\dagger_{l\sigma}
|7_g -2 \tau \rangle
\right].
\end{split}
\end{equation}
\end{widetext}
The sums over $l$ and $l'$ are a priori independent and run over all possible single particle angular momenta: $l,l' = -2,\ldots,3$. The angular momentum conservation of $H_{\rm m}$ implies, nevertheless, that the combinations which contribute to the sum must satisfy the condition

\begin{equation}
2-(-2) = l-l' \quad (\!\!\!\!\!\!\mod 6),
\end{equation}
which restricts the sum to the three pairs:
\begin{equation}
\label{eq:llp_cases}
\begin{cases}
 l = +2, \quad l'=-2;\\
 l = +1, \quad l'=+3;\\
 l = +3, \quad l'=-1.
\end{cases}
\end{equation}

Finally, it is not difficult to prove, starting from Eq.~\eqref{tunratetip}, the following properties for the elements of the rate matrix ${\bf \Gamma}^T$:

\begin{equation}
\label{eq:Gamma_symm}
\begin{split}
\Gamma^T_{ll'} &= |\Gamma^T_{ll'}|\e^{-\i(\phi_l-\phi_{l'})},\\
|\Gamma^T_{ll'}| &= |\Gamma^T_{\bar{l}l'}| = |\Gamma^T_{l\bar{l'}}|.
\end{split}
\end{equation}

Combining Eq.~\eqref{eq:H_eff_OffDiag_def} with  \eqref{eq:llp_cases} and \eqref{eq:Gamma_symm}, one obtains:

\begin{widetext}
\begin{equation}
\begin{split}
\langle 7_g \,2\,\tau | H_{\rm eff}^T |7_g -\!2\,\tau \rangle =&
\frac{1}{2\pi} \sum_\sigma
\left[
\langle 7_g \,2\,\tau |
d^\dagger_{2\sigma}
\left|\Gamma^T_{2,-2} (E_{7_g}-H_{\rm m})\right|\e^{-2\i\phi_2}
p_T(E_{7_g}-H_{\rm m})
d_{-2\sigma}
| 7_g -\!2\,\tau \rangle\right.\\
&+
\langle 7_g \,2\,\tau |
d_{-2\sigma}
\left|\Gamma^T_{2,-2} (H_{\rm m}-E_{7_g})\right|\e^{-2\i\phi_2}
p_T(H_{\rm m}-E_{7_g})
d^\dagger_{2\sigma}
| 7_g -\!2\,\tau \rangle\\
&+
\langle 7_g \,2\,\tau |
d^\dagger_{1\sigma}
\left|\Gamma^T_{1 3} (E_{7_g}-H_{\rm m})\right|\e^{-\i\phi_1}
p_T(E_{7_g}-H_{\rm m})
d_{3\sigma}
| 7_g -\!2\,\tau \rangle\\
&+
\langle 7_g \,2\,\tau |
d_{3\sigma}
\left|\Gamma^T_{1 3} (H_{\rm m}-E_{7_g})\right|\e^{-\i\phi_1}
p_T(H_{\rm m}-E_{7_g})
d^\dagger_{1\sigma}
| 7_g -\!2\,\tau \rangle\\
&+
\langle 7_g \,2\,\tau |
d^\dagger_{3\sigma}
\left|\Gamma^T_{3,-1} (E_{7_g}-H_{\rm m})\right|\e^{-\i\phi_1}
p_T(E_{7_g}-H_{\rm m})
d_{-1\sigma}
| 7_g -\!2\,\tau \rangle\\
&+
\left.\langle 7_g \,2\,\tau |
d_{-1\sigma}
\left|\Gamma^T_{3,-1} (H_{\rm m}-E_{7_g})\right|\e^{-\i\phi_1}
p_T(H_{\rm m}-E_{7_g})
d^\dagger_{3\sigma}
| 7_g -\!2\,\tau \rangle
\right],
\end{split}
\end{equation}
\end{widetext}
from which Eq.~\eqref{eq:H_eff_OffDiag} can be easily obtained.
It is now interesting to explore the different limits of  Eq.~\eqref{eq:H_eff_OffDiag}. In the constant interaction picture, for example, the term proportional to $B$ vanishes. The eigenstates of the interacting Hamiltonian $H_{\rm m}$ coincide in fact in the constant interaction model with the single Slater determinant eigenstates of the non interacting one. In practice, the 7 particle ground state $|7_g l \tau \rangle$ can be written as:
\begin{equation}
d^\dagger_{l\tau}|6_g\,0\,0\rangle,
 \end{equation}
with
\begin{equation}
|6_g\,0\,0\rangle = \prod_{l = -1}^{+1}\prod_{\tau = \uparrow,\downarrow} d^\dagger_{l\tau}|0\rangle.
\end{equation}
Thus, it follows immediately that:

\begin{equation}
\label{eq:CrAn7}
\begin{split}
d_{3\sigma}|7_g\,-\!2\,\tau\rangle &= 0,\\
d^\dagger_{1\sigma}|7_g\,-\!2\,\tau\rangle &= 0.
\end{split}
\end{equation}

By inserting Eq.~\eqref{eq:CrAn7} into the second equality in \eqref{eq:AB} one concludes that, in the constant interaction model,
the effective Hamiltonian for the $7$ particle ground state has the form:

\begin{equation}
\left(H_{\rm eff}\right)_{7g} = \left(
                                  \begin{array}{cc}
                                    K & A \e^{-2\i\phi_2} \\
                                    A\e^{+2\i\phi_2} & K \\
                                  \end{array}
                                \right),
\end{equation}
where the hermitianicity of the $H_{\rm eff}$ has been used. The constant $K$ obtained from the direct evaluation of Eq.~\eqref{eq:H_eff} is different from the off-diagonal constant A.
Nevertheless, any contribution to the $N, E, S_z$ subblock of the effective Hamiltonian which is proportional to the unity matrix does not influence the dynamics of the system (see Eq.~\eqref{GME}). Thus we chose to set $K = A$ which gives the form of the $H_{\rm eff}$ given by the Eqs. \eqref{omegaHeff} and \eqref{RotOP}. This choice is particularly interesting among all others since if $\theta_{\rm tip} = n\pi/3$ (for example when the tip is exactly above one of the carbon atoms) the operator $L$ defined in Eq.~\eqref{RotOP} is the generator of the discrete rotations around the axis passing through the center of the molecule and the carbon atom closest to the tip.

Finally let us consider under which conditions the effective Hamiltonian commutes with the stationary density matrix evaluated only taking into account the tunnelling dynamics. By combining Eqs.~\eqref{eq:H_eff_OffDiag} and \eqref{sigmastat} one eqsily obtains for the 7 particle ground state with spin $\tau$:

\begin{equation}
[H_{\rm eff},\sigma_{\rm stat}] = 2\i B_{H} B_{\sigma} \sin(2\phi_2 - \phi_1)\sigma_z,
\end{equation}
where $\sigma_z$ is the third Pauli matrix and we have introduced the subscripts $\sigma$ and $H$ to distinguish between the constants proceeding from the density matrix and the effective Hamiltonian.
In the constant interaction picture $B_H = 0$, while $B_\sigma = 0$ if the tip is respecting the rotational symmetry of the molecule, i.e. $\vec{R}_{\rm tip}$ is on the principal rotational axis of the molecule. Finally a vanishing condition can be obtained also from the phases when $2\phi_2 - \phi_1 = n\pi$. By assuming the approximate expression for the phase given by Eq.~\eqref{eq:Phase_approx} one gets $\theta_{\rm tip} = n\pi/3$ which corresponds to a tip belonging to one of the vertical mirror planes for the molecule intersecting a carbon atom. Notice that for these special values of $\theta_{\rm tip}$ Eq.~\eqref{eq:Phase_approx} is exact.

For completeness we conclude with the results regarding the 5 particle ground state. The effective Hamiltonian for the generic description of the Coulomb interaction reads:
\begin{widetext}
\begin{equation}
(H_{\rm eff})_{5_g \tau} = \left(
                             \begin{array}{cc}
                               K & A\e^{-2\i\phi_1} + B\e^{ -\i\phi_2} + C\e^{\i\phi_1}  \\
                               A\e^{-2\i\phi_1} + B\e^{ -\i\phi_2} + C\e^{\i\phi_1} & K \\
                             \end{array}
                           \right),
\end{equation}
where $A,\,B,\,C \in \mathbb{R}$ are given by

\begin{equation}
\label{eq:ABC}
\begin{split}
A =& \frac{1}{2\pi} \sum_\sigma
\left[
\langle 5_g \,1\,\tau |
d^\dagger_{1\sigma}
\left|\Gamma^T_{1,-1} (E_{5_g}-H_{\rm m})\right|
p_T(E_{5_g}-H_{\rm m})
d_{-1\sigma}
| 5_g -\!1\,\tau \rangle\right.\\
&\left.+
\langle 5_g \,1\,\tau |
d_{-1\sigma}
\left|\Gamma^T_{1,-1} (H_{\rm m}-E_{5_g})\right|
p_T(H_{\rm m}-E_{7_g})
d^\dagger_{1\sigma}
| 7_g -\!1\,\tau \rangle
\right],\\
B =&  \frac{1}{\pi} {\rm Re}\sum_\sigma
\left[
\langle 5_g \,1\,\tau |
d^\dagger_{2\sigma}
\left|\Gamma^T_{2 0} (E_{7_g}-H_{\rm m})\right|
p_T(E_{5_g}-H_{\rm m})
d_{2\sigma}
| 5_g -\!1\,\tau \rangle\right.\\
&\left.+
\langle 5_g \,1\,\tau |
d_{0\sigma}
\left|\Gamma^T_{2 0} (H_{\rm m}-E_{7_g})\right|
p_T(H_{\rm m}-E_{7_g})
d^\dagger_{2\sigma}
| 5_g -\!1\,\tau \rangle
\right],\\
C =&  \frac{1}{\pi} {\rm Re}\sum_\sigma
\left[
\langle 5_g \,1\,\tau |
d^\dagger_{3\sigma}
\left|\Gamma^T_{3 1} (E_{7_g}-H_{\rm m})\right|
p_T(E_{5_g}-H_{\rm m})
d_{1\sigma}
| 5_g -\!1\,\tau \rangle\right.\\
&\left.+
\langle 5_g \,1\,\tau |
d_{1\sigma}
\left|\Gamma^T_{3 1} (H_{\rm m}-E_{7_g})\right|
p_T(H_{\rm m}-E_{7_g})
d^\dagger_{3\sigma}
| 5_g -\!1\,\tau \rangle
\right].
\end{split}
\end{equation}
\end{widetext}

In close analogy with the 7 particle case, one proves that $B$ and $C$ vanish in the constant interaction picture and also that for $\theta_{\rm tip} = n\pi/3$ the effective Hamiltonian commutes with the stationary density matrix calculated only considering the tunnelling dynamics.

\end{appendix}

\end{document}